\documentclass[notitlepage,aps,prd,amsmath,floats,floatfix,
twocolumn,
superscriptaddress,nofootinbib,showpacs]{revtex4-1}

\usepackage{amsfonts,amsmath,units,wasysym,epsfig,graphicx,verbatim,color,subfigure,graphicx}
\usepackage{amsmath}
\usepackage{amssymb}
\usepackage{amsfonts}
\usepackage{bm}
\usepackage{color}
\usepackage[normalem]{ulem}
\usepackage{lipsum}
\usepackage{natbib}
\usepackage{color,soul}
\usepackage{hyperref}
\usepackage{marginnote}
\usepackage{braket}
\usepackage[none]{hyphenat}
\usepackage{bm}
\usepackage{comment}

\definecolor{green1}{RGB}{17,119,51}
\definecolor{yellow1}{RGB}{221,204,119}
\definecolor{pink1}{RGB}{204,102,119}
\definecolor{blue1}{RGB}{68,119,170}
\definecolor{blue2}{RGB}{136,204,238}

\begin{document}

\newcommand{\IUCAA}{\affiliation{Inter-University Centre for Astronomy and Astrophysics, Post Bag 4, Ganeshkhind, Pune 411 007, India}}
\newcommand{\IISERP}{\affiliation{Indian Institute of Science Education and Research Pune, Dr. Homi Bhabha Road, Pashan,  Pune  411008, India}}
\newcommand{\IISERK}{\affiliation{Indian Institute of Science Education and Research Kolkata, Mohanpur, West Bengal 741252, India}}
\newcommand{\WSU}{\affiliation{Department of Physics \& Astronomy, Washington State University,
1245 Webster, Pullman, WA 99164-2814, U.S.A}}
\newcommand{\SU}{\affiliation{Department of Physics, Syracuse University, Syracuse, New York 13244, USA}}


\title{Toward low-latency coincident precessing and coherent aligned-spin gravitational-wave searches of compact binary coalescences with particle swarm optimization}

\author{Varun Srivastava}\SU\IISERP
\author{K Rajesh Nayak}\IISERK
\author{Sukanta Bose}\IUCAA\WSU

\date{\today}


\begin{abstract}
We investigate the use of particle swarm optimization (PSO) algorithm for detection of gravitational-wave signals from compact binary coalescences. We show that the PSO is fast and effective in searching for gravitational wave signals. The PSO-based aligned-spin coincident multi-detector search recovers appreciably more gravitational-wave signals, for a signal-to-noise ratio (SNR) of 10, the PSO based aligned-spin search recovers approximately 26 $\%$ more events as compared to the template bank searches. The PSO-based aligned-spin coincident search uses 48k matched-filtering operations, and provides a better parameter estimation accuracy at the detection stage, as compared to the PyCBC template-bank search in LIGO's second observation run (O2) with 400k template points. We demonstrate an effective PSO-based precessing coincident search with 320k match-filtering operations per detector. We present results of an all-sky aligned-spin coherent search with 576k match-filtering operations per detector, for some examples of two-, three-, and four-detector networks constituting of the LIGO detectors in Hanford and Livingston, Virgo and KAGRA. Techniques for background estimation that are applicable to real data for PSO-based coincident and coherent searches are also presented.
\end{abstract}

\maketitle

\section{Introduction}\label{sec:intro}

The Advanced LIGO and Advanced Virgo detectors have now reached a sensitivity where they have already made successful observations of gravitational waves (GWs) from multiple compact binary coalescences (CBCs), such as several binary black holes (BBHs)~\cite{GW150914,GW151226,GW170104,GW170608,GW170814,GW150914PE} and a binary neutron star (BNS)~\cite{GW170817}. GWs from astrophysical source can provide great insight into fundamental physics ranging from testing the general theory of relativity \cite{berti2015testing}, validating no-hair theorem for black holes \cite{gossan2012bayesian}, constraining the NS equation of state (EOS) \cite{GW170817,de2018constraining,Bose:2017jvk}. In the upcoming years as the sensitivity of each detector improves one can probe deeper into the Universe. The ground-based detectors are capable of detecting GWs primarily from BBH, NSBH and BNS binaries, where the black holes masses are in the stellar to the intermediate mass range. We will study the use of particle swarm optimization (PSO) algorithm \cite{PSO1995}  detect these sources. 

The present GW searches, with PyCBC~\cite{Usman:2015kfa, PyCBCLive} and GstLAL~\cite{2017PhRvD..95d2001M}, rely on a template-bank approach to detect GWs from CBCs~\cite{Owen1996, owen1999matched}. The PyCBC template-bank~\cite{canton2017designing} for LIGO's second observation run (O2) comprises approximately 400k templates. Thus, for a given analysis segment of data from a detector, there are a total of 400k matched-filtering operations (MFOs). An early detection of a GW signal facilitates the follow-up of prospective sources in various regions of the electromagnetic (EM) spectrum promptly. The combined knowledge of EM emission spectrum and the gravitational wave signal can be used to put constraints on the NS equation of state~\cite{EOSGW170817, de2018constraining,Margalit_NSEOS_2017}. The electromagnetic counterparts are GRBs, kilonovae and various other transient and longer-lived signals arising from BNS and in some cases NS-BH mergers. The kilonovae are expected to evolve over a timescale from a few hours to a few days \cite{Metzger2016}. These light curve predictions were verified by the electromagnetic observations of GW170817 \cite{GW170817Spectra}. This time constraint demands for prompt GW alerts for EM follow-up. The PSO algorithm \cite{PSO1995} provides a fast way to search for optimal solutions in a given parameter space. This property can be exploited to speed up the process of detection of GWs. 

In our work, we demonstrate the effectiveness of the PSO algorithm in GW data-analysis by successfully executing an all-sky blind coherent search that uses a total of 576k MFOs per detector. We also show that the PSO can effectively search for precessing binaries using coincident statistics with a total of 320k MFOs per detector. The high-speed and effectiveness of the PSO algorithm to provide optimal solutions in a given parameter space makes the above two searches feasible. PSO can also be used to speed up the online aligned-spin CBC search for GW signal in data. 

The template-bank approach \cite{owen1999matched} for detecting the GWs from CBCs in LIGO-VIRGO detectors is discretized (analogous to setting up grid-points in parameter space). The template bank is generated by choosing a minimal overlap/match \footnote{The {\em match} here is a measure  of the cross-correlation between the data and a known template, and is defined more precisely in Sec. \S\ref{sec:dgws}.} of $M = 0.97$ between the templates generated from two neighboring grid points in parameter space \cite{owen1999matched}. 

The template bank is generated prior to doing a search. The template bank generation is dependent on the geometry of the search parameters, which are used in its construction. Non-stationarity of detector noise can affect the geometry of template placement and, thus, require reconstruction of the bank. The generation of template banks is computationally expensive. The template bank for the second LIGO science run (O2) \cite{canton2017designing} consisted of approximately 400k templates. Though several efficient template placement algorithms exist, they become computationally expensive if the search parameter has to be extended owing to the increasing number of parameters or their ranges. However, in the PSO, the waveforms are computed on the fly and do not require a bank of templates. Moreover, the PSO algorithm actively learns during the search and places points using the information gathered during the search, making the search more optimal than the template bank.

We develop PSO and apply it in simulated examples to overcome some problems like the generation of template banks and a transition from discretized search algorithm to continuous search methods. In the future, our method can also be adapted to speed up the estimation of CBC parameters; however, there are some interesting challenges to be addressed. The match filtering operation itself is a computationally expensive task. PSO is a stochastic global optimization algorithm that quickly provides optimal solutions to a given function over its parameter space. This property can be exploited not only to reduce the computational cost of the search but also to reduce the errors in the estimation of CBC parameters in the detection stage. The application of PSO in GW data-analysis was proposed by Wang and Mohanty \cite{WangMohanty2010}. The performance enhancement and the use of PSO to set up a blind all-sky coherent search was demonstrated in Ref.~\cite{mohanty2017performance}. 

In our work, we use over 25k injections with different CBC parameters to characterize the PSO search, in contrast to Ref.~\cite{mohanty2017performance} which uses a handful of injections in a smaller parameter space of CBCs. In contrast to previous studies, we setup robust background estimation techniques for PSO based searches using time-slides which can be readily used when working with real data. We use the reduced cost of PSO searches to setup a full precessing coincident search for CBCs. We also show that PSO can be used to develop an all-sky blind coherent search pipeline. The challenge in the latter is to do coherent background estimation, which is computationally expensive. The PSO algorithm also provides more flexibility as the geometry of the parameter space is automatically manifested by the algorithm. The points for template generation are produced dynamically. The search parameters, the template approximants can be actively changed and the swarm intelligent algorithm will evolve accordingly to generate the search template points, giving greater flexibility. We will discuss these in greater detail in sections \S\ref{sec:PSO} and \S\ref{sec:res}. We will first describe briefly the aspects to gravitational wave data analysis. We will use both the coherent search and coincident search statistics to detect gravitational waves from CBCs, we will briefly review the idea behind these searches in section \S\ref{sec:dgws}. The PSO algorithm and the different variants explored in our study are discussed in section \S\ref{sec:PSO}.  We will present the results of our study in section \S\ref{sec:res}.

\section{Particle Swarm Optimization} \label{sec:PSO}
Particle swarm optimization (PSO) \cite{PSO1995} originated with the aim to model the social behavior of animals living in a colony like swarms. However, simulations reveal that the eventually swarms converged to optimal solutions of a given function ({\it fitness function}) in the parameter space. 

Being social gives that individuals in the swarm the ability to communicate with each other. This grants each individual in a given swarm or colony to be aware of the information gathered by all the other individuals in the swarm. The trajectory taken by an individual particle can then be modeled by three independent behavioral traits. These factors influence the path taken up by an individual during the period when it explores the parameter space. The first factor is {\it inertia}, implying that if an individual is moving in a given direction it will continue to move in that direction. The second trait is related to {\it nostalgia}, at any given instance of time, each individual has the tendency to give up the search and move towards the best location explored by it, referred to as personal best (pBest). The last factor is based on the collective ({\it social}) pool of knowledge. As each individual is aware of the best location explored by the entire swarm, the global best (gBest), each individual has a tendency to give up the energy exhaustive search process and move towards the global best location. These three factors influencing the velocity of an individual can be combined as
\begin{equation}\label{eq:psoVevo}
\begin{split}
v_{i,d}(t+1) = &~\omega \cdot v_{i,d}(t) + \gamma_p \cdot r_p \cdot (p_{i,d}(t) - x_{i,d}(t) ) \\
& + \gamma_g \cdot r_g \cdot (g_d(t) - x_{i,d}(t))
\end{split}
\end{equation}
where the subscript {\it i} marks the individual of the swarm, {\it d} represents the dimension. The factors $r_{p}$ and $r_{g}$ are stochastic factors which determine at any given instance of time the instinct to move towards the pBest and gBest respectively. The coefficients $\omega$, $ \gamma_{p}$ and $\gamma_{g}$ determine the strength of the inertial, nostalgic and social factors respectively. The force of attraction towards the global and personal best is assumed to be due to a harmonic potential. The above PSO notations are consistent throughout the paper. The position evolution equation is simply given by
\begin{equation}\label{eq:psoXevo}
x_{i,d}(t+1) = x_{i,d}(t) + v_{i,d}(t)
\end{equation}
Thus, using PSO one can obtain optimal solutions of a given {\it fitness function} in a parameter space. The steps of the PSO algorithm are briefly defined below.
\begin{itemize}
    \item Let $\mathcal{S}$ be the given parameter space in {\it D}-dimension. A swarm of $N_{p}$ particles is initialized with the particle's position vector $x_d$ having a uniform prior distribution in each dimension. Each particle is also assigned with a random velocity vector, $v_d$ with a norm uniformly distributed in the range $[-1, 1]$. The initialized pBest of each particle is the same as its initial position vector. The initialized global best is computed by evaluating the fitness function $f(x_d)$ for the initial location of each particle, and then the maximum value of $f(x_d)$ achieved is used to initialize the gBest location.
    \item The swarm is evolved with the equations \ref{eq:psoVevo} and \ref{eq:psoXevo}, in discrete time steps of one. The pBest and gBest locations are updated by checking if the fitness function evaluated at the new position vectors is higher than the previous step.
    \item The evolution is terminated with a stop condition. Usually, a maximum number of steps $N_{s}$ are predefined and used as a flag to stop the simulation. The value of gBest location at the end of the simulation is considered as one of the optimal solutions to the fitness function $ f(x_d)$.
\end{itemize}
We use the PSO algorithm to maximize the likelihood function and find optimal solutions in the given parameter space of compact coalescence binaries(CBCs).

We extended the PSO scheme to search using multiple swarms over the parameter space. There are two ways this can be achieved. One in which all the swarms evolve independently of each other and thus, can be evolved simultaneously. The second way is to evolve subsequent swarms in a way that they use the information acquired by previous swarms ({\it relayed}). The latter can be used to give priors on search variables (instead of a subsequent blind search) and reduce the extent of parameter space. The subsequent swarms, in this case, can use the information acquired by the searches carried out by previous swarms. The former search method has the advantage that it can be easily parallelized across machine nodes.

\begin{figure}
\begin{center}
\includegraphics[width=0.49\textwidth]{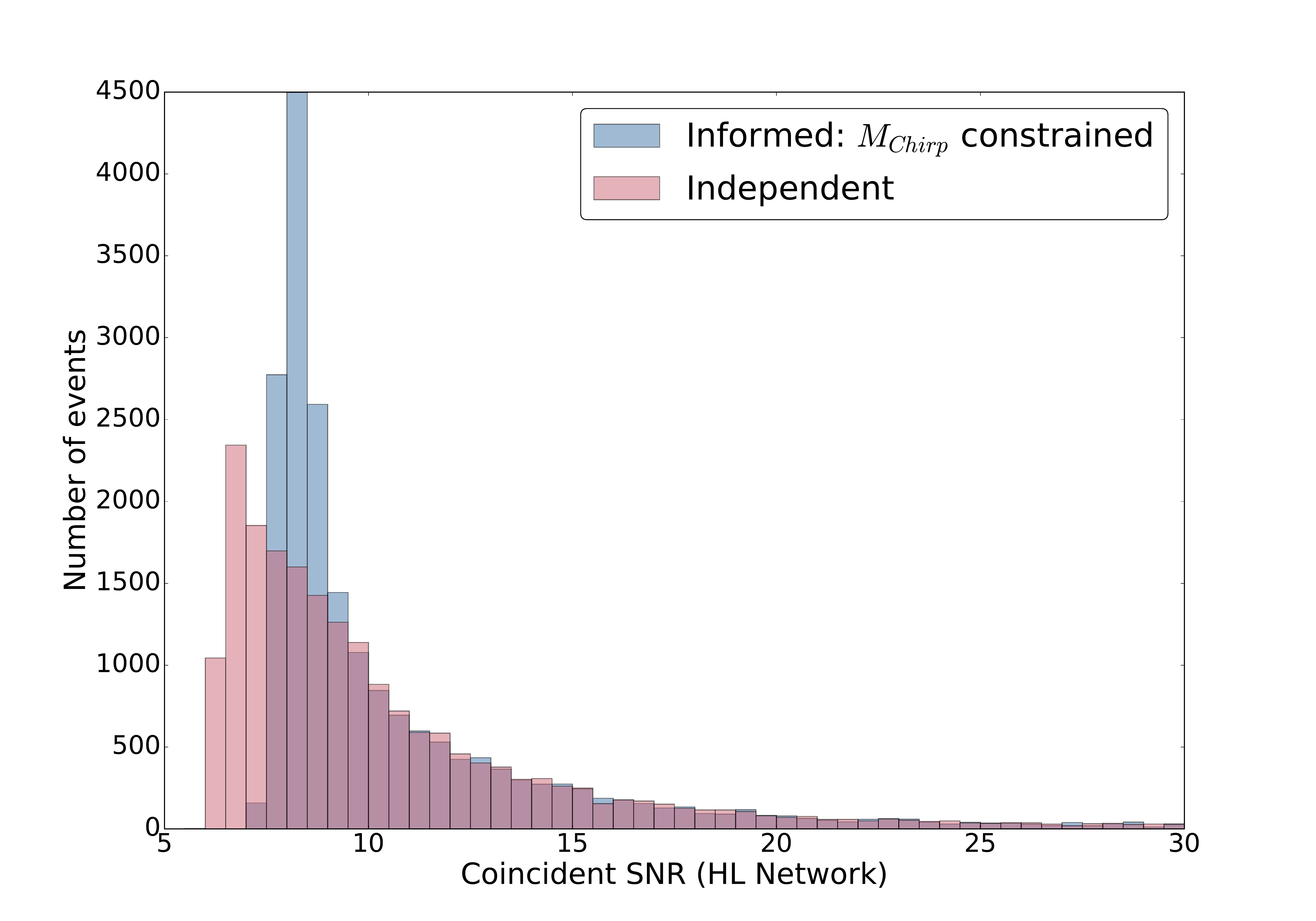}
\caption{The figure compares the distribution of the SNR of the two sub-swarms in the case where -- the second sub-swarm evolves independently of the first (pink) and the case where the second swarm uses the $\pm$ 30 $\%$ error on $M_{Chirp}$ obtained by the first swarm. For 20k injections, we see that the later scheme performs better as the sub-swarms achieve an overall higher signal to noise ratio (SNR).}
\label{fig:MchComp}
\end{center}
\end{figure}

We use the information from the multiple swarm variants of the relayed searches in two ways. The first way is using the gBest information of previous swarms and using the hostile swarm algorithm discussed in section \ref{subsec:HPSO}. The second method is using the information about the parameters estimated. In GW data analysis, chirp mass ($M_{chirp}$) is one of the parameters that is recovered with good accuracy. We can use the $M_{chirp}$ estimated from previous swarms to set up a uniform prior on the subsequent swarms. Figure \ref{fig:MchComp} compares the performance between a search where multiple swarms are independent to the one with the $M_{Chirp}$ constraint. We see that having the $M_{Chirp}$ prior significantly improves performance. Next, we describe some of the alternatives to the standard PSO algorithms to enhance the performance specific to our problem. 

\subsection{Multiple Independent Swarms} \label{subsec:MultPSO}

The simplest extension to the standard PSO algorithm is the use of multiple swarms independent of each other to explore the parameter space. The use of multiple swarms is advantageous in the situation when the parameter space has multiple optimum solutions. The evolution equations of velocity and position are the same as \ref{eq:psoVevo} and \ref{eq:psoXevo} for any given sub-swarm. At the end of the simulations, one obtains an optimal solution corresponding to each sub-swarm. From this set of global best location explored by each sub-swarm, we choose the one which gives the highest value of the fitness function -- i.e. matched filter or likelihood as the case may be. 

\subsection{Hostile Swarm Algorithm} \label{subsec:HPSO}
Many species in nature compete amongst themselves and exhibit territorial behaviors. We aim to extend and modify the PSO algorithm to model the territorial behavior. In nature animal species maintain territories because of limitation of resources. Individual within a certain group interact amongst themselves but have a hostile attitude towards any intruder. Thus, the multiple sub-swarms are no longer independent but have a hostile attitude towards each other. The velocity equation for a given sub-swarm is thus modified to 
\begin{equation}\label{eq:hpsoVevo}
\begin{split}
v^{m}_{i,d}(t+1) = &~\omega \cdot v^{m}_{i,d}(t) + \gamma_p \cdot r_p \cdot (p^{m}_{i,d}(t) - x^{m}_{i,d}(t) ) \\
& + \gamma_g \cdot r_g \cdot (g^{m}_d(t) - x^{m}_{i,d}(t)) \\
& - \sum_{n;{n \ne m}} \gamma_r \cdot r_r \cdot \mathcal{F}(g^{n}_d(t) - x^{m}_{i,d}(t))
\end{split}
\end{equation}
where the superscripts {\it m} (or n) represent the m-th sub-swarm. The function $\mathcal{F}$ represents the repulsion force by which the hostile sub-swarms interact with each other. The functional chosen for the repulsion can be arbitrary. We use an inverse square repulsion potential. However, when deciding the functional form of repulsion it is important to consider the size of the parameter space and the strength of repulsion. If the repulsion force is set to be very high then particle's in a swarm will leak out of the parameter space and the boundary condition will randomly re-inject the particle back in the parameter space. If there are a large number of particles leaving the parameter space at each step, then the swarm intelligence will not evolve and optimal solutions will not be rendered. One way to step around this problem is re-normalizing the particle's velocity to $v_{norm}$ if the particle's velocity is greater than some $v_o$.

The particles in a given swarm are repelled at any instant of time from the gBest location discovered by any other sub-swarms allowing an effective exploration in the parameters space as not only the convergence is delayed but the repulsion leads the sub-swarms to search unexplored locations in parameter space. However, there are some more caveats to be addressed for the method to function effectively. One major problem arises if multiple hostile swarms are evolved simultaneously. The problem arises that by chance if any of individuals in two or more distinct swarm end up by chance exploring some optimal (global or local) solution during the course of their exploration. In an extreme case, consider two swarms end up with the same gBest location then their convergence is drastically affected due to the repulsion. Multiple hostile swarms search can be relayed to solve these issues.

Sometimes, in a relayed search as well, the second swarm despite the repulsion may end up close to the optimal solution previously explored and from where it is repelled from again causing convergence issues. To further reduce such the probability of this to happen one can further add a distance constrained. To elaborate, if them ({\it m = 1, 2, ... }) evolved swarms have explored gBest locations $\vec{g}_{1}, \vec{g}_{2}, ...$. Then the m+1 swarm is evolved in a way that it is constrained that none of the individuals in the swarm can have their pBest or gBest closer than the distances ${d}_{1}, {d}_{2}, ...$ gBest of them previous swarms. Lastly, because of the repulsion potential, the convergence of the swarms may be adversely affected. When the repelled swarm has evolved then to allow swarm convergence we turn off the repulsion and evolve the repelled swarm with the standard PSO evolution equation for five steps. This ensures that the swarm smoothly converges. The velocity of each individual is also reinitialized to very small values so that they don't fly away from the converged location. All hostile swarm searches end with five iterations using the standard PSO evolution equations. 

\subsection{Remove PSO} \label{subsec:rmPSO}
In GW detection problem, we wish to generate possible triggers of detected astrophysical events as soon as possible. The PSO algorithm provides very quick swarm convergence if there is a very sharp outlying global peak. This property could be exploited in GW data analysis. The following trick helps reduce the computational cost, specially in cases of GW triggers with higher SNRs. In general, the following modification is specially relevant in cases where the cost of computation of the fitness function is large. In the last few iterations of the PSO search, the particles wander around the gBest with slight or no improvement in the estimate of gBest. For the purpose of detecting the gBest location, sampling the peak and calculating the posterior is not important, thus the late iterations of PSO could be avoided if the swarm has converged reducing the computational cost. We allow the swarms to evolve for 40 iterations, however, if the SNR of the trigger is considerably high the swarm converges early. Consider N iterations were allowed for the swarm to evolve, we choose to remove the particles that have converged after M ($<$ N) iterations. We can remove the particles based on the following criteria and ways. First, consider a hypersphere of radius r in the dimensional space around the global best location, then removing a fraction f of the particles within this volume at every iteration following M. One could also consider a hypercube around the global best location and remove particles within it in the same way. The second way to eliminate the grouped particles is by dividing the dimensions $d_1, d_2,...$ into $n_1, n_2,...$ bins respectively. If there are multiple particles in any hypercuboid in the parameter space after M iterations, we remove a fraction f of the particles in that hypercuboid, each step following M.

\subsection{PSO Variants Summary}
The performance of different PSO variants was compared. We use eight sub-swarms in our maximization and find that multiple-independent swarms and the different variant of hostile swarms all performed comparably. We use multiple-independent swarm for the rest of the study. In our subsequent sections, all the results are presented for this variant. 

\section{Gravitational Wave Data Analysis} \label{sec:dgws}
The matched filter is the technique prominently used to meaningfully extract gravitational wave signals buried in detector noise \cite{owen1999matched}. Given some data imprinted with signal from an astrophysical source along with noise \footnote{Data can be decomposed as $s^{I}(t)$ = $n^{I}(t)$ (noise) + $h_{T}^{I}(t,\xi)$ (astrophysical signal)}. The matched-filter function is defined as the cross-correlation of templates\footnote{$h^{I}(t,\xi)$ represents a time-domain template generated with parameters $\xi$ and data is represented by $s^{I}(t)$.} $h^{I}(t,\xi)$ (of known parameters) with detector ($I$) data. The matched filter is weighted by the power spectral density of noise in frequency domain which is a representative of detector sensitivity across the frequency band. The detector noise $n^{I}(t)$ is related to the power spectral density of noise $S^{I}_{h}(f)$ by the fourier space auto-correlation function of noise $\tilde{n}^{I}(t)$ given by
\begin{equation}
    \braket{\tilde{n}^{I}(f)|\tilde{n}^{I}(f')} = \delta(f-f')S^{I}_{h}(f)
\end{equation}
For a single detector the matched filter \footnote{The fourier transform of a time-series a(t) is denoted by $\tilde{a}(f)$.} is given by equation \ref{eq:mf1exp}.  
\begin{equation} \label{eq:mf1exp}
    (s^{I}|h^{I}) = 4 \mathcal{R}e \int_{0}^{\infty} df~\frac{ {\tilde{h}^{I*}(f)\tilde{s}^{I}(f)} }{ S_{h}^{I}(f) }\,.
\end{equation}
For a network of detectors, the matched filter function is simply the sum of matched filter computed for a given template in all the detectors in that network. In the absence of an astrophysical signal, the detector output is a time series of noise. Detecting gravitational waves implies distinguishing between the former case against the presence of astrophysical signals in data. To distinguish between the two one can either use Bayesian inference approach or a frequentist approach. In Bayesian inference, the likelihood ratio is defined as the ratio of the probability that data has signal present (Test Hypothesis) to the probability of no signal being present (Null Hypothesis), mathematically expressed as:
\begin{equation}
    \lambda(h) = \frac{P(s|h)}{P(s|0)} = \frac{ e^{-(s^{I}-h^{I}|s^{I}-h^{I})\cdot0.5} }{ e^{-(s^{I}|s^{I})\cdot0.5} }
\end{equation}
The above expression translates to the following log-likelihood measure for a given detector
\begin{equation} \label{eq:1Dlike}
    \ln\lambda = (s|h) - \frac{1}{2}(h|h)
\end{equation}
Based on the parameterization of the templates used for matched filtering, one can describe two different statistical approaches -- coincident and coherent. We will address the {\it coherent} and {\it coincident} GW search methods in detail for multi-detector scenarios in section \S\ref{sec:coin} and \S\ref{sec:coher} respectively. We briefly describe the dimensionality and extent of the parameter space of CBCs in section \S\ref{subsec:CBCPS}. Next, we will describe the injection parameters used in our study along with the parameters of PSO search in section \S\ref{subsec:InjParam}

\subsection{Parameter Space of Compact Binaries}\label{subsec:CBCPS}
The GW waveform from CBCs depends on seventeen parameters, when including generically spinning components and eccentric orbits. Here we will consider non-spinning as well as spinning components in various simulation studies, but never eccentric orbits. While a long inspiral into LIGO-like sensitivity band is expected to render the orbit devoid of any eccentricity, certain evolutionary scenarios allow for in-band (late-time) residual eccentricities that cannot be ignored. Here we choose to set aside the study of eccentric orbits for a future work. Thus, the signals studied here will require atmost 15 parameters. These parameters can be divided into two categories -- intrinsic and extrinsic. The intrinsic parameters are inherent to the source, such as the masses of the stellar object $M_{1}$ and $M_{2}$ in a binary and their respective spin vectors $\bm{S_{1}}$ and $\bm{S_{2}}$. On the other hand, the extrinsic parameters are the source luminosity distance $D$, the inclination of the orbital plane of the binary $\iota$ with respect to the line of sight, the source sky-position angles $(\theta, \phi)$, the polarization angle $\psi$, the coalescence phase $\phi_{o}$, and the time of arrival $t_{a}$ of the signal at a given detector location. The matched-filter function (eq. \ref{eq:mf1exp}) involves cross-correlation of tens to a thousand of seconds long data segment, making a search in a fifteen-dimensional parameter space computationally expensive. 

The search statistics employed to detect GWs in the network of detectors can be divided into two -- the coincident and the coherent search. Each of these searches is tuned such that the search space is mathematically reduced and there is a subset of parameters $p_s$ over which the search is carried. Some of the other parameters can be estimated given the estimates of individual parameter $p_s$. Further in many GW searches, the search space is reduced to lower dimensions in order to reduce this computational burden and to allow the possibility to make prompt GW detections, crucial to EM follow-up. Typically the effects of spins is reduced by considering the component spins of the individual objects in the binaries to be either aligned or anti-aligned with the orbital angular momentum. As a result, the six-dimensional parameter extent of spin is reduced to two ($S_{1z}$, $S_{2z}$). For LIGO-O2 the bank incorporates for aligned spins $S_{1z}$ and $S_{2z}$ and component mass parameters \cite{canton2017designing}. The total number of template points in the PyCBC-O2 template bank were 400k. 

\subsection{Injections and PSO Parameters} \label{subsec:InjParam}
The aLIGO detectors are sensitive enough to detect GWs from CBCs in a frequency range from close to 20 Hz to a few thousand Hertz. For effectively utilizing the computational time, we divide the parameter space of CBCs into different groups. 
\begin{itemize}
    \item BBH Injections: The masses of individual components in this group range from 12 $\textup{M}_\odot$ to 80 $\textup{M}_\odot$. We divide the BBH injections into sets -- high and low spin BBH injections. The spins on component masses are aligned or anti-aligned with the CBCs angular momentum. Each subset consists of 10k different injections. In the sub-group of low spin BBH CBCs, we constrain the spins of individual objects to be in the range from -0.5 to 0.5. For the high-spin BBH injection set one of the component masses is forced to have the spin in range (0.5 to 0.85), whereas the second object can either be aligned or anti-aligned with the former with a spin up to 0.85. The injections are distributed uniformly in SNR over the volume with minimum coincident SNR of 5.5 and the minimum SNR in any given detector of 4.
  \item BNS Injections: The masses of individual NS is in the range from 1 $\textup{M}_\odot$ to 2 $\textup{M}_\odot$. We also assume that the BNS systems will have low effective spins when they are close to the merger \cite{SpinConstNS}. Thus, the spins of each individual object go up to a maximum of 0.05. We allow the possibility of individual spins to be aligned or anti-aligned. We generate 500 BNS injections distributed uniformly in SNR over the volume with minimum coincident SNR of 7.5 and the minimum SNR in any given detector of 5. The increased minimum is to allow the recovery of a greater number of events and overcome the small number statistics of recovered events.
  
\item Non-precessing NS-BH Injections: The masses are for this injection set are so chosen that the injections are more likely to give rise to electromagnetic counter parts~\cite{GBM:2017lvd}. The black hole mass is restricted to range from 5 $\textup{M}_\odot$ to 14 $\textup{M}_\odot$. The NS mass and spins are varied in the same range as before. We restrict the spin on black holes to a maximum of 0.4. We have assumed that the spin of binaries is aligned. For higher black-hole spins the assumption breaks, the coupling of component spins with the orbital angular momentum will cause precession \cite{PrecessingNSBH}. We generate 2k injections smeared uniformly in SNR over the volume with minimum coincident SNR of 7.5 and a minimum SNR in any given detector of 5.
  \item Precessing NS-BH Injections: The masses and SNR distribution range is same as that of non-precessing NS-BH injection set. The difference is we allow the possibility of precessing spins in NS-BH binaries. The total BH spin is smeared uniformly in the range of 0.5 to 0.85. We generate 2k injections in this set.
\end{itemize}

We use the theoretical design power spectral density (PSD) of aLIGO detectors~\cite{Shoemaker2009}, VIRGO and KAGRA to generate noise in each detector using PyCBC \cite{pycbc-software}. The lowermost sensitive frequency of each detector is assumed to be 20 Hz for the generation of templates. The injected signals and the templates generated are sampled at 4096 Hz. We use IMRPhenomD 3.5PN \cite{IMRPhenomD} templates for PSO based aligned-spin coincident and coherent searches. For precessing injections and template points, we use IMRPhenomPv2 waveform model \cite{IMRPhenomPv2_2014,IMRPhenomPv2_2015}. For the template bank search similarly, the injections and the search uses SEOBNRv4-ROM-DoubleSpin waveform model \cite{SEOBNRv4} for a total mass greater than 4~$M_{\odot}$, which is in accordance with the template bank \cite{canton2017designing}. We ensure that the injected signal and the waveform model used in the search are the same. The approached opted for in our study is the following. The GW signal from the astrophysical parameters discussed above is simulated. Then, we add noise to the simulated GW signal, the noise is whitened by the PSD of the corresponding detector. Then, the swarms are initialized and are used to optimize for the coherent SNR or the coincident SNR. 

In aligned-spin coincident search, we use 40 iterations of PSO and eight multiple swarms with 150 particles each to explore the parameter space, a total of 48k MFOs per detector. In precessing-spin coincident search, we use 40 iterations of PSO and 10 multiple swarms with 800 particles each to explore the parameter space, a total of 320k MFOs per detector. The PSO parameters in equation \ref{eq:psoVevo} are set to $\omega$= 0.5,   $\gamma_{p}$ = 2 and $\gamma_{g}$ = 2. We use an electrostatic repulsion potential and a linear repulsion potential to repel the particles of subsequent swarms from the gBest of previous in hostile swarm algorithm. The parameters in \ref{eq:hpsoVevo} are the same values along with $\gamma_{r(lin)}$ = 1.05 $\gamma_{g}$ and $\gamma_{r(ES)}$ = 25. The different variants of PSO had similar performance. All our results are presented for multiple-independent swarm variant of PSO.

\section{Coincident Search} \label{sec:coin}
The ground-based GW detectors have an antenna-like all-sky sensitivity, lacking the ability to locate the GW source in the sky. Using multiple detectors and triangulation techniques the source location is determined. Additionally, environmental or instrumental disturbances give rise to glitches in the detector. Some of these glitches mimic the GW signal from CBCs \cite{DetChar_O1, DetChar_GW150914}. One fundamental discriminator to veto these glitches is the time of arrival in a network of GW detectors with similar sensitivity operating at the same time. The astrophysical signals in the detectors cannot be separated in time by a time greater than the light travel time $t_c$ between any two corresponding detectors in the network. The glitches are uncorrelated across detectors and this approach drastically reduces the false positives arising due to the glitches. All our injections are into gaussian noise. Another discriminator of noise glitches and signal is the chi-squared discriminator \cite{AllenChisq}.

In a network of detectors the coincident SNR is computed by match filtering a given template $\tilde{h}^{I}( \xi, f)$ with data in each detector $\tilde{s}^{I}(f)$. However, the match-filter output from detectors has to be combined keeping the light travel time distance constraint across any two detectors. To compute the coincident SNR, one of the detectors is taken as a reference detector and the matched filter series is computed. The maximum of the match-filter gives the time of arrival at the reference detector $t_a^{ref}$. Next, the match-filter output for the same template is computed across all other detectors. Given a light travel time between the reference detector and some detector $j$ in the network $t_c^{ref;j}$. In coincident search, for $j$ detectors  the maximum of match-filter output in the time window  $t_a^{ref} \pm 1.5 t_c^{ref;j}$, gives the time of arrival $t_a^{j}$ in the $j$ detector. The value of the maximum of match-filter at  $t_a^{j}$ is added in quadrature to give the coincident SNR for the reference detector. The reference detector is then changed and the coincident SNR is recomputed. During the search, the template's parameters $\xi$ are varied over the search space. We define the best template which maximizes the coincident SNR with template parameters $\xi = \xi_{max}$. If the coincident SNR for template $\xi_{max}$ is greater than some thresholds defined to discriminate against noise and signals of astrophysical origin, we flag the event of astrophysical importance. The parameters $\xi_{max}$ are the first-hand estimates of source parameters. Using algorithms like nested sampling, one can calculate the posterior and estimate parameters with greater accuracy and better sampling in an offline search which is not constrained by time.

We use the PSO algorithm to maximize over the coincident SNR described above. The template parameters we maximize our search over are $M_{chirp}$, $\eta$, $s_{1z}$, $s_{2z}$ and $\iota$ (inclination of the orbit), in the general case. Thus, the dimensionality of the search space is five. We include the orbital inclination as an independent parameter as in the upcoming years with multiple GW detectors the distance-inclination degeneracy is expected to be broken \cite{DIDegen}. To show that the PSO search is not drastically affected by the change is search parameters, we will compare the search with $m_{1}$, $m_{2}$, $s_{1z}$, $s_{2z}$ and $\iota$ as search parameters in section \S\ref{subsec:coin_param_flex}. For consistency check we will compare the performance of PSO search over high-spin and low-spin BBH injections in section \S\ref{subsec:coin_cbcs}. We also summarize the results of PSO based aligned-spin coincident search on BNS and NSBH injections, in section \S\ref{subsec:coin_cbcs}. Next, we will use PSO to set up a precessing coincident search (dimensionality of nine). The results of the precessing coincident search over precessing NS-BH injections and the corresponding comparison with PSO based aligned-spin coincident search over the same set of injections is summarized in section \S\ref{subsec:Coin-Pres}. In section \S\ref{subsec:coin_bg} we will estimate the background of PSO based coincident and coherent searches using 100k gaussian noise realization and 30 time-slide over each of them. Lastly in section \S\ref{subsec:comp-O2-Coin} we will compare the performance of PSO based coincident search with the O2-template bank. We will then vary the number of detectors in the network and present the performance of PSO in section \S\ref{subsec:coin_ndet}.

To summarize the errors in the estimation of different CBC parameters in different searches we will use box-and-whisker plots throughout the paper. The box and whiskers plot is a projection of a histogram. The limits of the colored box extend from the lower quartile to upper quartile. ``Whiskers'' plotted on either side of the box extend to 1.5$\times$ inter-quartile range (IQR) \footnote{If the histograms were Gaussian, the ends of the whiskers would be at  4.7$\sigma$ on either side of the mean.}. Outliers are points outside the whiskers and are marked as `+' signs.
 
\subsection{PSO based aligned-spin coincident searches over BBH, BNS and NSBH injections}\label{subsec:coin_cbcs}
\begin{figure*}
\begin{center}
\includegraphics[width=0.99\textwidth]{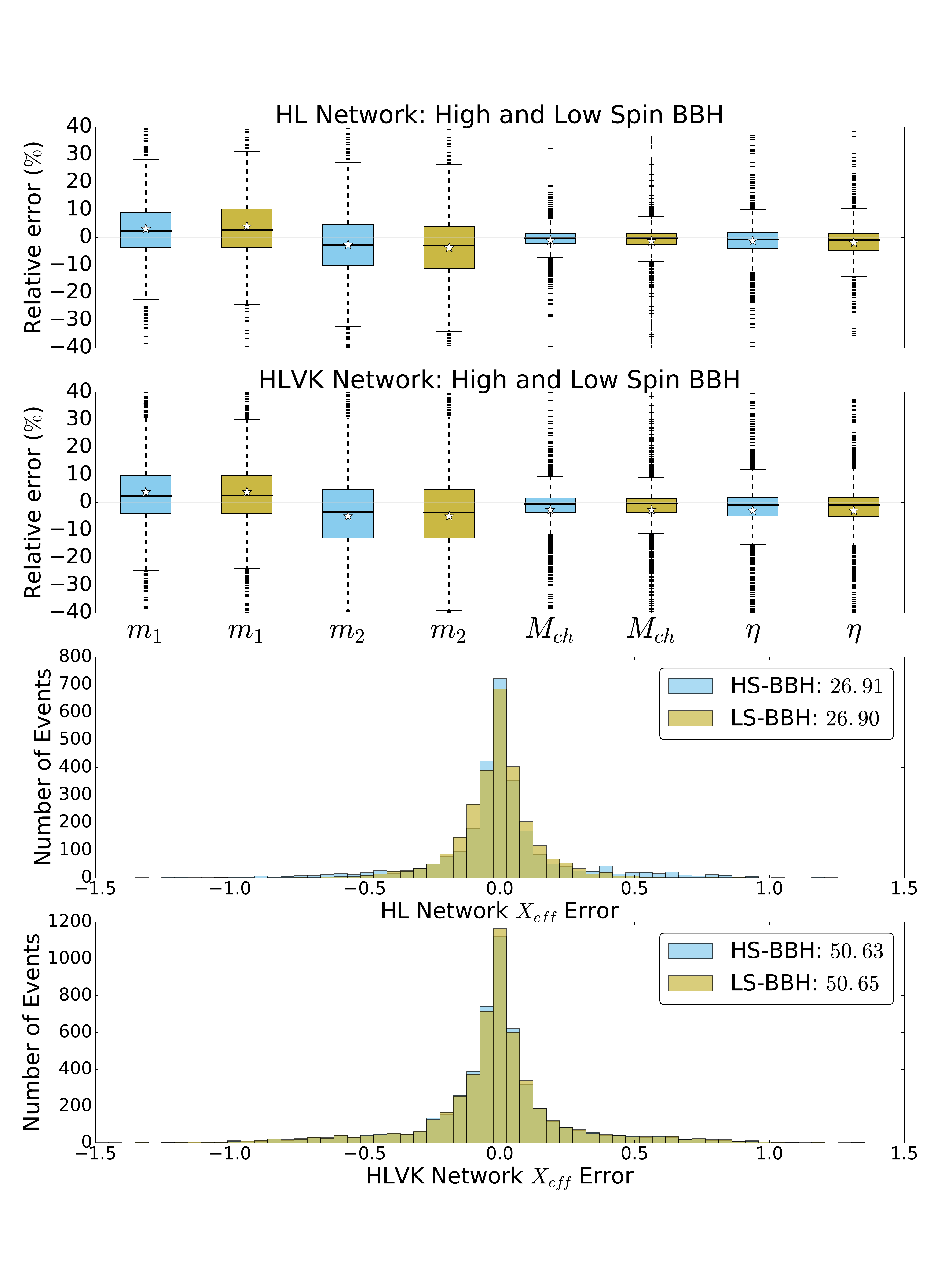}
\vspace{-10mm}
\caption{The plot above summarizes the errors on the different parameters of binary black-hole injections. For a total of 10000 injections in each set, the errors shown above are for events that pass the corresponding network thresholds. We see for both the detector network configurations, the coincident SNR maximization using PSO is effective to recover signals from both the injection sets -- high-spin (Blue) and low-spin (yellow). The error distribution is almost comparable for the HL and HLVK network as a majority of the events in HLVK are recovered using two detectors in the network, this is also the reason for the slight under-performance.}
\label{fig:coin_spin_comp}
\end{center}
\end{figure*}

To estimate the performance of PSO algorithm and to test the effectiveness of dynamic generation of points in parameter space, we test our algorithm by comparing against the different set of injections defined in section \S\ref{subsec:InjParam}. First, we test the performance of PSO to recover BBH injections. The BBH injections are divided into two categories -- high and low spinning. Each of the two injection sets contains 10k injections, smeared uniformly component masses, but the total spin of the two system varies as per the corresponding definitions in section \S\ref{subsec:InjParam}. We use the PSO algorithm to search over the parameter space and maximize the coincident SNR for an HL and HLVK detector network. The observations are summarized in Fig.~\ref{fig:coin_spin_comp}. Placing a threshold of 10 and 14.25 on the HL and HLVK detector networks (explained in section \S\ref{subsec:coin_bg}), we see the fraction of events recovered from the two different parts of the parameter space of BBH are almost the same and the error in estimation of parameters also has a similar distribution. 

\begin{figure}
\begin{center}
\includegraphics[width=0.49\textwidth]{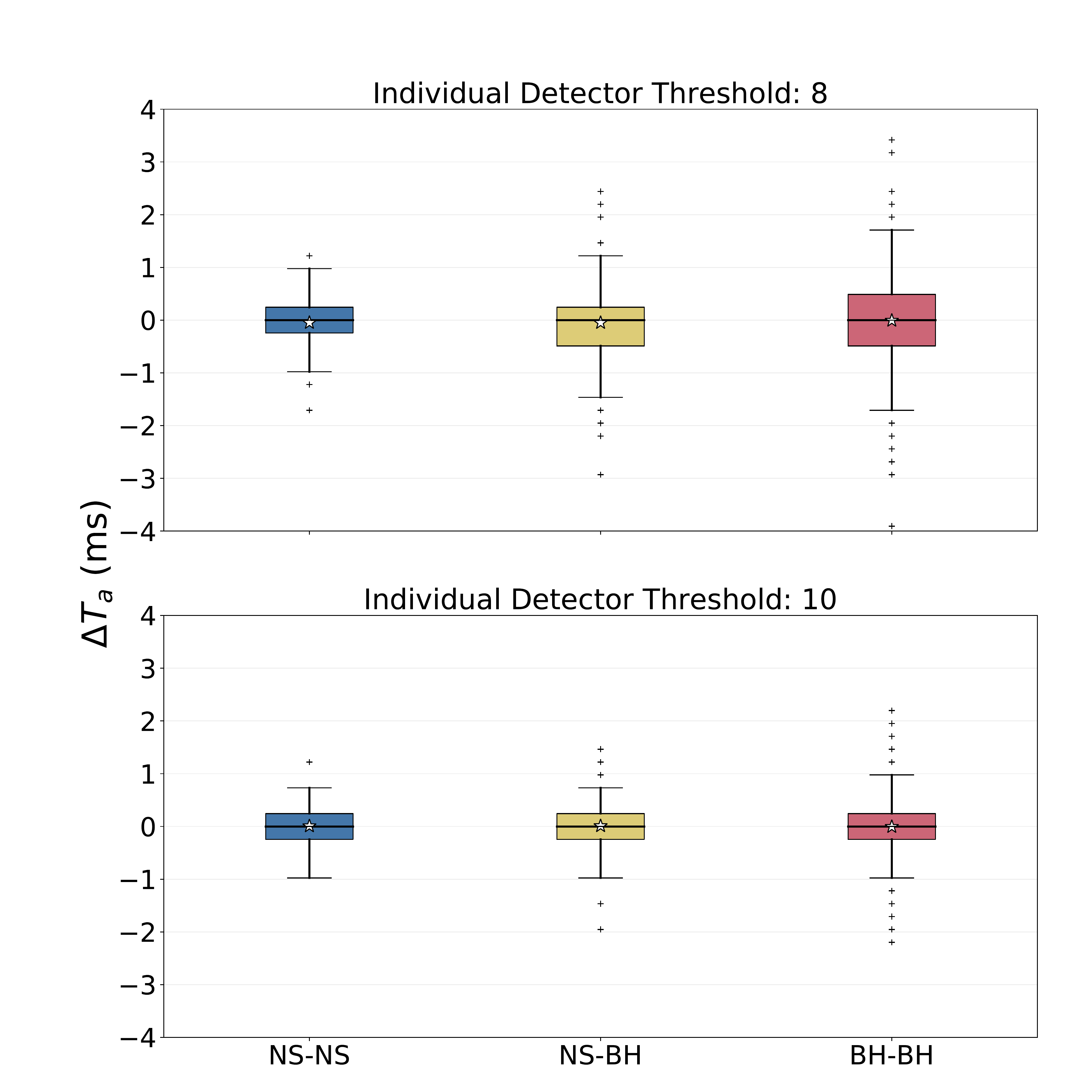}
\caption{For the different injection sets -- BBH, BNS, and NSBH (defined in section \S\ref{subsec:InjParam}) used in our study the above plot summarizes the error in the time of arrival in a network of two detectors (HL) for the injections that were recovered. The thresholds on individual detectors used to flag an injection as recovered are shown above the corresponding subplot. From the figure above it is evident that the time of arrival is effectively measured in an aligned-spin coincident search.}
\label{fig:2D_TA}
\end{center}
\end{figure}

\begin{figure*}
\begin{center}
\includegraphics[width=0.99\textwidth]{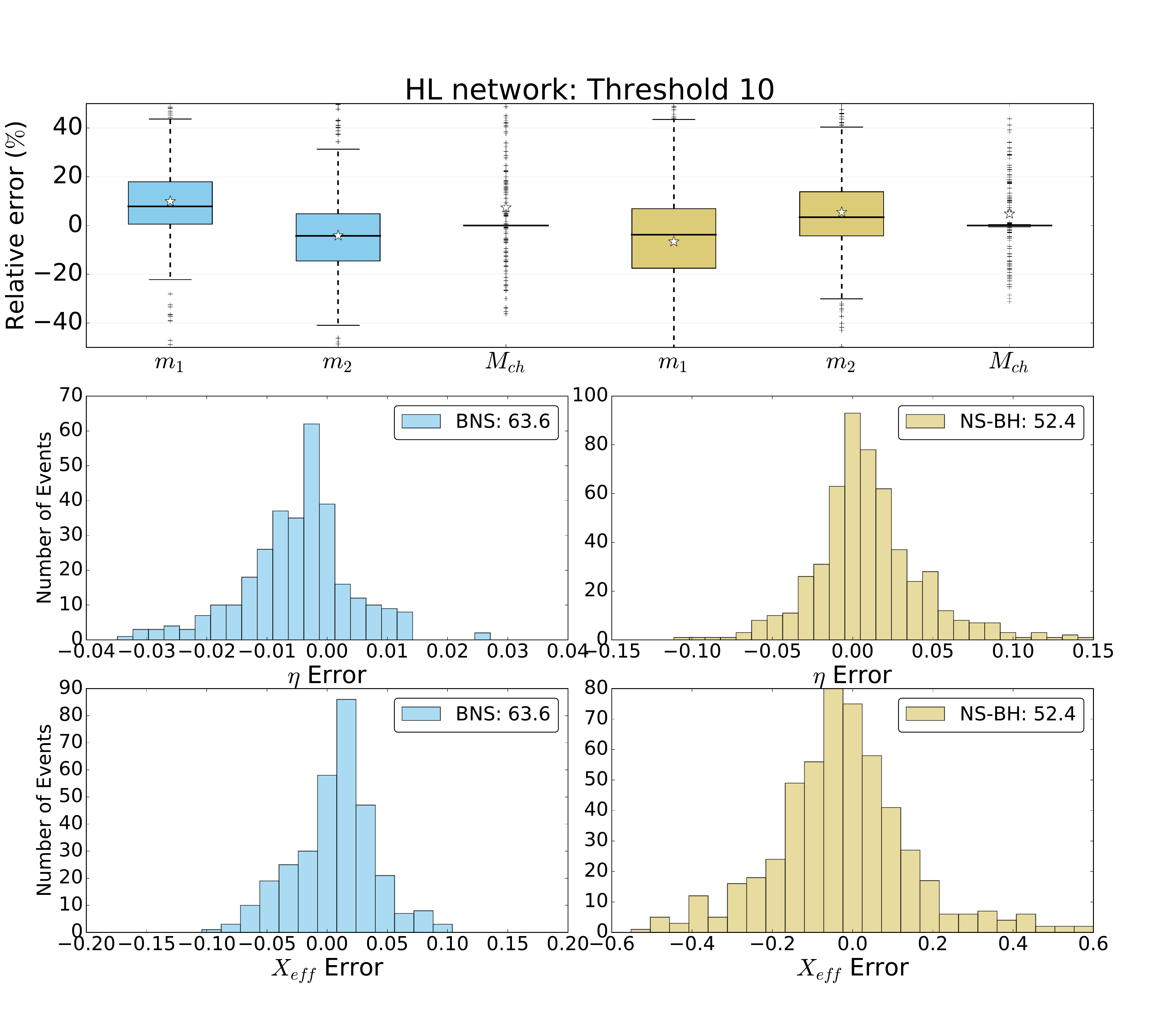}
\caption{The plot summarizes the relative errors in the estimation of parameters for 500 BNS (blue) and 2k NSBH (yellow) injections. From the plots above we see that the $M_{Chirp}$ is estimated with a great accuracy, a trademark characteristic of BNS and NSBH binaries. The estimated errors on other parameters are also consistent with the expected values from the statistical approach described in \cite{ajith2009estimating}.}
\label{fig:BHNS-BNS-2Det}
\end{center}
\end{figure*}

We will now extend the PSO method to a search over aligned spin BNS and NSBH injections. This combined with the above two searches completes the parameter space of the CBCs that can be detected by ground-based GW detectors. We do a similar exercise as discussed before and maximize the coincident SNR for these injections in HL detector network. The results are summarized in Fig.~\ref{fig:BHNS-BNS-2Det}. We get a precise measurement of the $M_{chirp}$ value which is a trademark characteristic feature of binaries with NS. The error in other parameters is also lesser compared to BBH search. Lastly, the Fig.~\ref{fig:2D_TA} summaries the error in the time of arrival in milliseconds for the different injection sets. Thus, we have demonstrated that the dynamic template placing in PSO algorithm is effective to recover signals, almost independent of the component spin of objects and the nature of CBC.

\begin{figure}
\begin{center}
\includegraphics[width=0.49\textwidth]{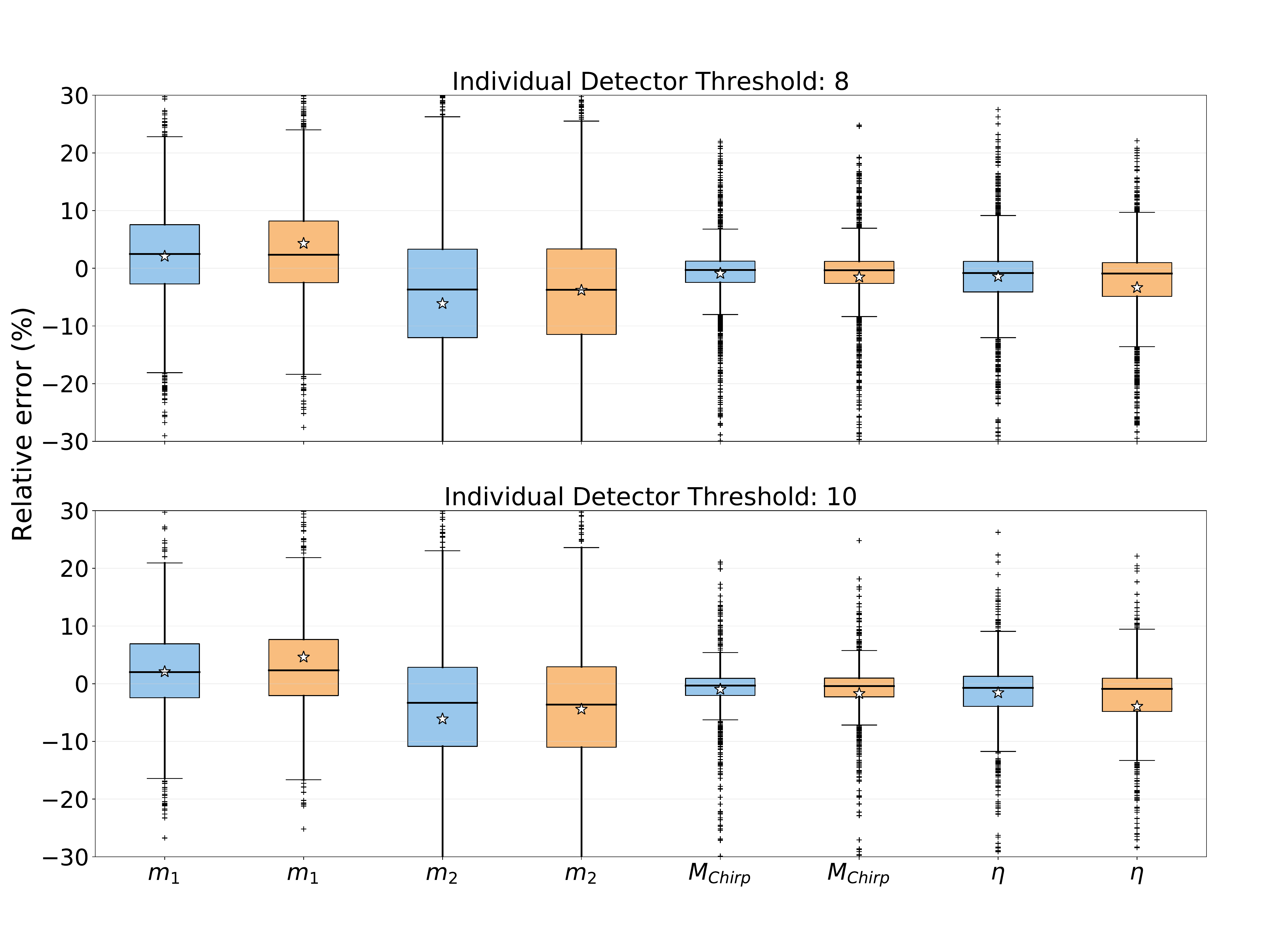}
\vspace{-10mm}
\caption{ For 20k BBH injections, the plot above summarizes the results when the PSO based coincident search with two detectors(HL). A search with mass parameters $(m_1,m_2)$ (blue) -- the masses of individual objects in binaries is compared against a search done with mass parameters $(M_{chirp}, \eta)$ (orange) -- the chirp mass and the symmetric mass ratio of the binary. The plot shows that the performance is almost identical in the two search methods. The number of events passing the thresholds (labeled above each subplot) in each detector is also equal.}
\label{fig:MchEtaSrch}
\end{center}
\end{figure}

\subsection{Flexibility of Search}\label{subsec:coin_param_flex}

The functional form of the ambiguity function changes with the choice of parameters used in the search~\cite{Dhurandhar:2017aan}. Specifically, how the match of a given template with its neighboring templates falls off with increasing difference in their parameter values varies with the parameter choice, such as ($m_1, m_2$) as opposed to ($M_{chirp}, \eta$). The performance of a template-bank based search depends on such choices~\cite{Owen1996}. The question that arises is whether such choices affect PSO-based searches as well. We use the same 20k BBH injections described in the section above to look for any such effects. 

We set up two different searches over the 20k BBH injections simulated in the HL detector network. One with parameters ($m_1, m_2, s_{1z}, s_{2z}, \iota$) while the other with parameters ($M_{chirp}, \eta, s_{1z}, s_{2z}, \iota$). Figure \ref{fig:MchEtaSrch} compares the result of the two searches. We see that the results, in terms of detection efficiency and error estimates are almost similar. The rate of convergence to the optimal solutions might be different amongst the different families of search parameters but for reasonable iteration steps, PSO search indicates weak dependence on the search parameters allowing more flexibility.

\begin{figure*}
\begin{center}
\includegraphics[width=0.99\textwidth]{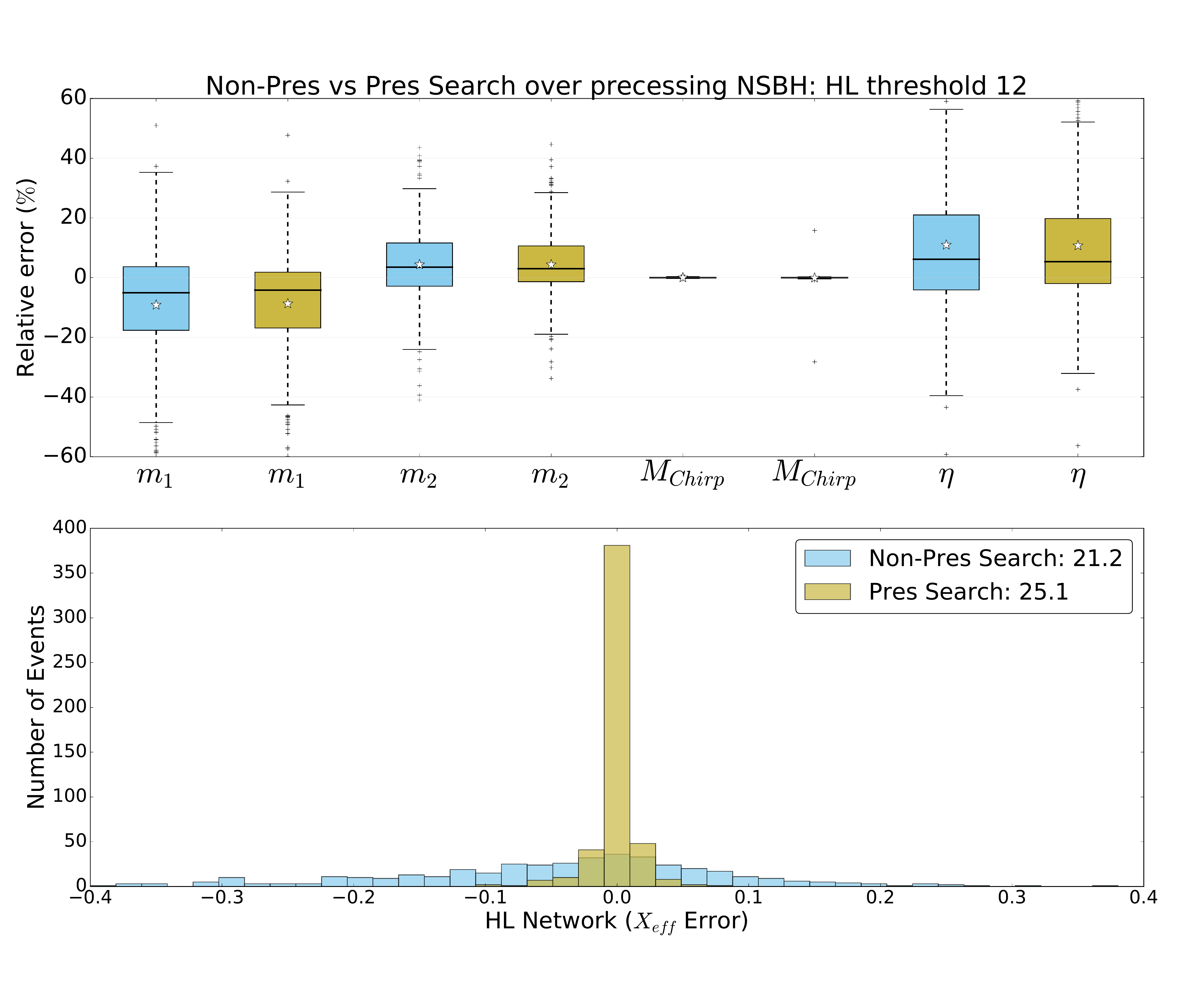}
\caption{For the 2k precessing NSBH injection, the plot compares the performance of a precessing PSO search (yellow) with an aligned-spin PSO search (blue). The background is similar for the two, but for the same false alarm probability, the detection efficiency of precessing search is higher. We see for the same set of injections with identical noise realization, precessing search recovers approximately 18 $\%$ more events compared to an aligned-spin PSO search with the same number of MFOs. The error in mass parameters has similar distribution for the two searches but the estimation of $\chi_{eff}$ is very accurate in a precessing PSO search. }
\label{fig:Pres_NSBH}
\end{center}
\end{figure*}

\subsection{Precessing NSBH search} \label{subsec:Coin-Pres}
In astrophysical scenarios, in NSBH binaries the BH are expected to have high spins due to the accretion of the NS matter onto the BH, this makes it more likely for the orbit to precess \cite{PrecessingNSBH}. We aim to extend the CBC search from aligned spin to precessing search using PSO to search for precessing NSBH.

We first define the parameters of the injected signal in this sub-domain of precessing NSBH signals. We generate 2k injections where the mass of BHs range from 5 $\textup{M}_\odot$  to 14 $\textup{M}_\odot$ and their total spin range from 0 to 0.85. For the NSs the mass ranges between 1 $\textup{M}_\odot$ to 2 $\textup{M}_\odot$ whereas their total spin is restricted in the range from 0 to 0.05. These values are chosen from astrophysical knowledge of these systems such that their GW emission is detectable by the ground-based detectors.

To recover these signals we set up a nine-dimensional search over parameters ($m_1$, $m_2$, $S_1$, $S_2$, $\iota$). Using PSO we maximize the coincident SNR. However, to ensure that the search is effective in this high-dimensional parameter space we increase the number of particles in the PSO algorithm to 800. The number of independent swarms used is also increased to 10. The total number of 320k MFO are performed in the process. We aim to check the capabilities of PSO to perform a precessing search to recover precessing injections. To look for any performance improvements or advantages obtained from a precessing search, we compare the results of the search over precessing NSBH injections with precessing PSO search against aligned-spin search ($m_1$, $m_2$, $S_{1z}$, $S_{2z}$, $\iota$), with identical PSO parameters, swarm-size, and number. Thus, the total number of MFOs are same and equal to 320k in both the searches. 

The results of the two searches are summarized in Fig.~\ref{fig:Pres_NSBH}. Using the same threshold on the two searches, we see out of the 2k precessing NSBH injections, the precessing search recovers 25.1 $\%$ compared to a recovery of 21.2$\%$ from aligned-spin PSO search. It has been demonstrated that a non-precessing search would underperform with respect to a precessing search when searching over precessing injections \cite{IanPres2014}. By extending our search to account for precession we improve the recovery of injections by almost 18 $\%$ with respect to PSO aligned-spin coincident search. The second striking feature of the precessing search is the accuracy in the estimation of the $\chi_{eff}$ parameter of the precessing binary. In summary, PSO based precessing search is promising and with parallelization techniques could be used in an online search with 320k MFOs. 

\begin{table*}
  	\begin{center}
  	\begin{tabular}{|c|c|c|c|c|c|}
  		\hline
        \ Threshold & Corresponding & Coincident & Coincident & Coherent & Coherent \\
        \ ({\color{blue}Coincident}, {\color{red} Coherent})& False Alarm Probability & TB & (PSO based) & No-spin & Spin  \\
        \hline \hline
		\ {\color{blue} 8.5, \color{red} 8} &  $ 5 \times 10^{-7} $ & 76.9 $\%$ & 70.4 $\%$ & 88.4 $\%$ & 87.5  $\%$\\
        \hline
        \ {\color{blue} 9, \color{red} 8.5} &  $ < 10^{-7}$ & 44.6 $\%$ & 46.6 $\%$ & 56.8 $\%$ & 54.1$\%$ \\
        \hline
        \ {\color{blue} 10, \color{red} 9.5} & $ < 10^{-7}$ & 21.0 $\%$ & 26.9 $\%$ & 32.4 $\%$ & 29.6$\%$ \\
        \hline
  	\end{tabular}
  	\end{center}
    \caption{We use 100k gaussian noise realizations and 30 time-slides on each to estimate the background. Using time slides over gaussian noise injections, we estimate the false alarm probability (FAP) from the figure \ref{fig:BG-2Det} for PSO based coincident and coherent searches. Next, from the background estimation plots we define the thresholds for PSO based coincident and coherent searches. The above table summarizes the fraction of events recovered in different searches for the same false alarm probability in the background, see Fig.~\ref{fig:BG-2Det}. We see that for a non-degenerate network of detectors HL, the PSO based coincident search recovers a higher number of events from the same injection set as compared to the template bank search. For a coincident SNR threshold of 9, the PSO based coincident search recovers approximately 4.5$\%$ more events compared to template bank search. For a SNR threshold of 10, the PSO based coincident search outperforms the template bank coincident search by 26$\%$. Thus, the dynamic template placing capabilities of PSO help reduce the SNR loss by achieving a higher match. The injected signals also have anti-aligned spin injections, which makes the effect more stark \cite{canton2017designing}. We use the estimated coherent background to define thresholds on coherent search. Using the 20k BBH injections in HL network and comparing the performance of coincident searches with coherent searches at the same FAP, we find that coherent search recovers much higher number of events than coincident search. This is due to the sharp drop in the coherent background which is observed and expected \cite{IanHarry2010,Ian2016} in a case of non-degenerate detector network HL. }
    \label{tab:BG_Th}
\end{table*}

\subsection{Background Estimation}\label{subsec:coin_bg}
\begin{figure}
\begin{center}
\includegraphics[width=0.5\textwidth]{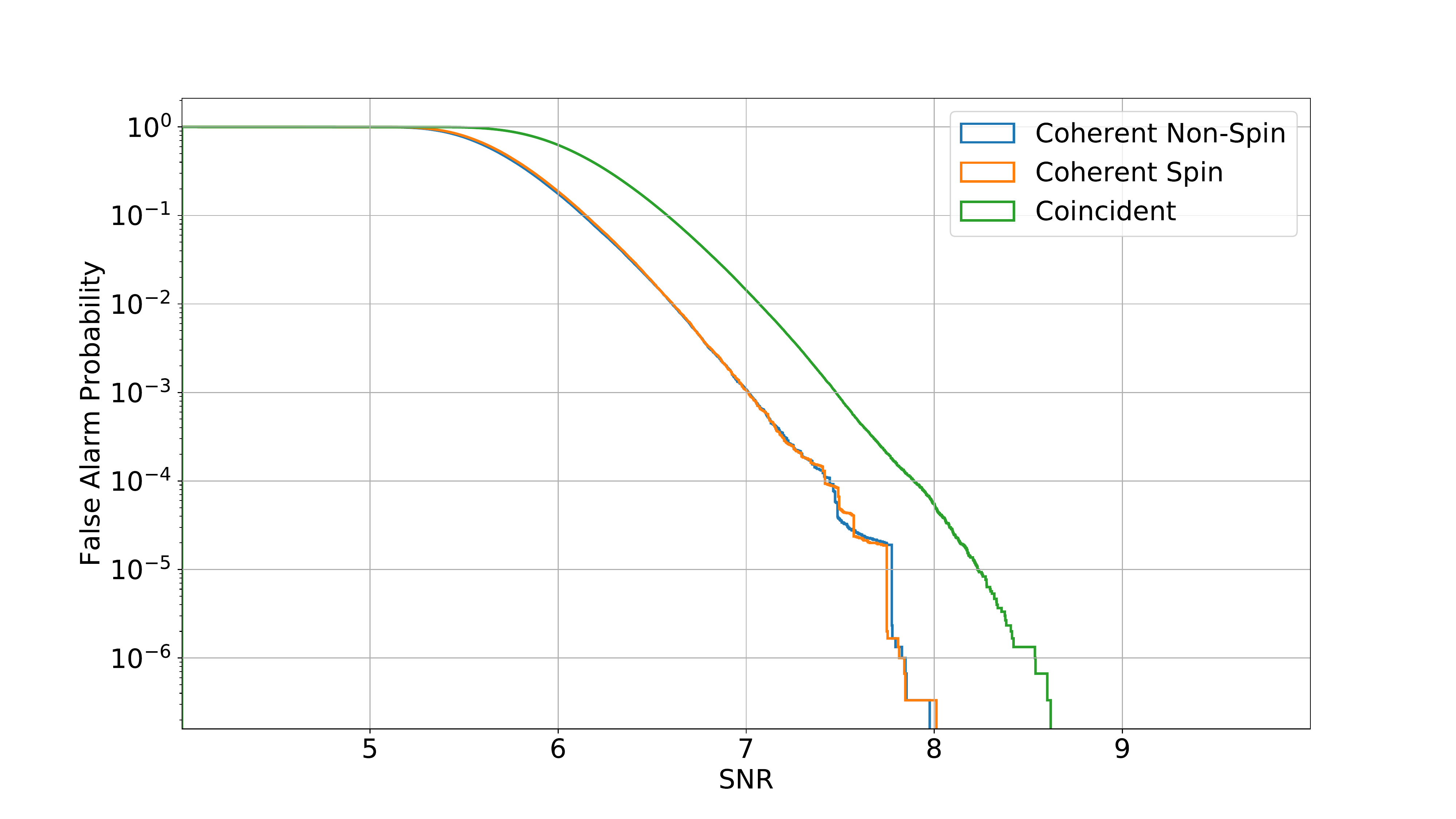}
\caption{We take 100k noise realizations and estimate the background by time sliding on each of the noise realizations. We estimate the background for coincident search, non-spinning coherent search and spinning coherent search using the method described in section \S\ref{subsec:coin_bg}. The plot above summarizes the background for each search in a network of two detectors (HL). For a false alert probability of the order $\mathcal{O}(10^{-6})$, the network coincident SNR is close to 8.5, for the same false alarm probability the coherent searches have an SNR close to 8. The sharp drop in coherent SNR arises due to and is a trend expected for a non-degenerate detector network \cite{IanHarry2010}. We point out that the three statistics are different and their corresponding SNR values alone for any injection are not the true measure of their effectiveness. For a given injection, our coherent SNR will be substantially higher than the coincident SNR. Thus, for performance comparison, we note that while for a false alarm of unity, the coincident SNR is $\sim 8.5$ and the coherent SNRs are $\sim 8$ many more injections are recovered in the latter case than the former case at that false alarm value. }
\label{fig:BG-2Det}
\end{center}
\end{figure}

We use 100k gaussian noise injections to estimate the background of the coherent and coincident PSO based searches. To estimate the background of a given gaussian noise stream, we use PSO to maximize the corresponding SNR -- coincident or coherent. The background events are not correlated. For a coincident background trigger, the time of arrival between the two detectors of the network must be greater than the light travel distance between the two detectors -- no astrophysical signal will be separated in two detectors greater than the light travel time. To compute the background we use time-slides on the dynamic template points which generate an SNR greater than 5 in one detector. To compute the background coincident SNR we take a 200ms time window in the stream of other data ensuring that this time window doesn't overlap with the corresponding time of arrival window for astrophysical signals in the former detector. The maximum SNR in the 200ms time window of the second detector in added in quadrature with the SNR of the first detector. The background estimation of the coherent search in done in the same way, only the time window of the coherent astrophysical signal is removed from the time-slide window. We do 30 time-slides for each noise realization. The Fig.~\ref{fig:BG-2Det} summarizes the estimated background for the two searches in a network of two detectors. From the figure, we estimate the false alarm probability of each search statistics. 

We use the thresholds corresponding to the respective false alarm probability over the 20k BBH injections used in our study. The fraction of these injections which cross the thresholds is flagged as recovered events. The fraction of events recovered from the 20k BBH injections, each with a different but unique noise realization consistent with different search methods, are summarized in the table \ref{tab:BG_Th}. The coherent search is computationally more expensive than the coincident searches. However, the fraction of events recovered by the coherent search is higher than the fraction of events recovered by coincident search. The PSO based coincident search also recovers more events with higher SNRs compared to the template bank used in out study, for the same set of injections due to the dynamic placing of template points and its ability to find optimal solutions of a function in parameter space.

\begin{figure*}
\begin{center}
\includegraphics[width=0.99\textwidth]{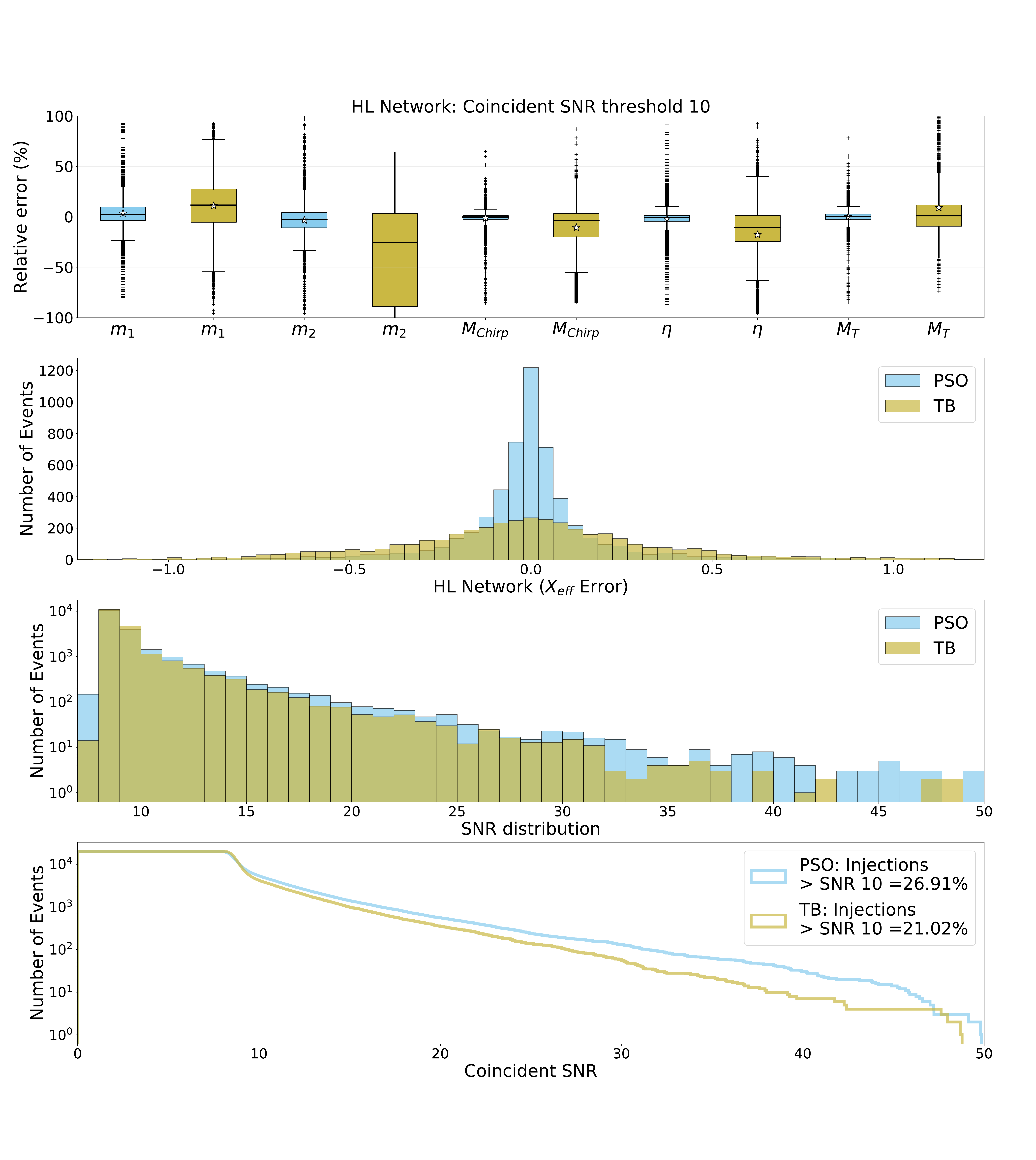}
\vspace{-5mm}
\caption{Comparison between template bank (TB, shown in Yellow) and PSO (Blue) to search over 20k BBH injections. PSO uses a total of 48k matched-filter operations whereas the O2-template bank has 400k template points. The first two plots from the top compare the error of coincident signal recovered with an SNR greater than 10. We see that PSO estimates parameters of the binaries with less error as compared to the TB search. The third plot shows the recovered SNR distribution of both the searches for all the injections. The last plot shows the SNR distribution for events recovered with an SNR greater than 10. The last plot shows that PSO performs better for high SNR injections with approximately 28$\%$ more events than template bank recovered with SNR greater than 10 -- the same injection set and with an identical noise in each injection set in each detector in the two searches. For an SNR threshold of 9, we find that the PSO based aligned-spin spin search recovers approximately 4.5$\%$ more events as compared to the template bank. This improvement obtained by PSO is due to the algorithm's capability to place template points dynamically and find the optimal solution in the parameter space to maximize the coincident SNR. The O2 template bank has higher SNR loss for anti-aligned spins \cite{canton2017designing}. A majority of the 28 $\%$ improvement in SNR with PSO arises in the SNR range from 10 to 14. However, for lower SNR events PSO under-performs in the SNR window from 7-9. This is partly due to the drifting of events to higher SNRs using PSO, as the total number of events in both the searches is the same. The third plot shows PSO recovers fewer events in SNR range of 7-8 than TB. This effect is mitigated by increasing the matched-filter operations in PSO to 96k.}
\label{fig:Comp_O2}
\end{center}
\end{figure*}

\subsection{Comparison with O2-Template Bank}\label{subsec:comp-O2-Coin}
We compare the performance of the PSO based coincident search with the O2-template bank over the 20k BBH injections. The noise realization of each injection in each detector is unique and identical for both the searches. The template bank searches rely on the discretization of the parameter space of CBCs in a way that the points in the bank have an overlap of 97$\%$ with the neighboring templates in the bank. To detect GW signals the strain data is match-filtered with all the points in the template bank. For multiple detectors, the data is combined using coincidence statistics. For the template bank search, the threshold value of SNR for a given injection in a single detector is 4.5. The new-SNR threshold for the template bank search has a single detector threshold of 4. In the template bank and PSO based searches, we use the same sampling rate of 4kHz and the noise realization for any given injection is identical in the corresponding detectors. We combine the triggers from the template bank search and get the template that maximizes the coincident SNR for the given injections.

Figure \ref{fig:Comp_O2} compares the template bank and PSO based aligned-spin searches in an HL detector network. The PSO based search uses 48k MFO only. We see PSO performs better except for low coincident SNR range of 7-8 where template bank has fewer number of events. Implying that points in the template bank yielded higher SNRs whereas PSO based search resulted in lower SNRs. The performance of the PSO algorithm improves when particles are increased by a double, 96k MFOs. As evident from Fig.~\ref{fig:Comp_O2}, the estimation of parameters is much more efficient with PSO. The $\chi_{eff}$ is much more accurately measured in a PSO based search. The errors on parameters estimated are consistent with the statistical errors predicted in \cite{ajith2009estimating}. If we consider the fraction of events from the 20k injections that are recovered with an SNR greater than 10, we see PSO based search has a higher fraction. PSO outperforms template bank search by recovering approximately 28 $\%$ more events with SNR greater than 10. For an SNR threshold of 9, the PSO based coincident search recovers approximately 4.5$\%$ more events than the template bank search. Higher recovered SNRs in the detection process also help reduce the error in sky-localization using BAYESTAR \cite{Bayestar2016}.

\section{Coherent Search}\label{sec:coher}
In the coincident search, described in the previous section, the constraint on data to discriminate from astrophysical signal was that of time of arrival difference between corresponding detectors. However, by virtue of its origin, gravitational waves are emitted coherently. Thus, GWs have an additional property of being coherent across each detector. If we put this constraint that the signals are coherently emitted from a source at location ($ \theta, \phi$). Thus, if the data has any signal of astrophysical origin it too would be coherent across multiple detectors which noise would not. By choosing a source at location ($ \theta, \phi$), one can find the time delay between the detectors as GWs travel at the speed of light. If the source location is known then the search is a targeted coherent search. If the source location is not known then the search is blind coherent search. In blind coherent search, the source location is a variable. By choosing different locations across the sky as a parameter along with other intrinsic parameters of CBCs, the coherent matched function can be maximized using PSO. We use the dominant polarization basis to get the coherent wave statistics in our search using PSO. For a detailed discussion on coherent search  refer to \citep{IanHarry2010,BoseCoher2011}. We will present a brief summary of coherent statistics. 

Using the same definitions of variables and functions defined in section \ref{sec:dgws}. The GW signal can be broken down into two polarizations $h_{+}$ and $h_{\times}$. Each of the polarization can be expressed in phase and amplitude terms, dependent on the response of the detector \cite{IanHarry2010}. For a given template parameter the gravitational waveform in the $I^{th}$ detector is given by

\begin{equation}
	h^{I}(t) = \sum_{\mu = 1}^{4} A^{\mu} h_{\mu}^{I}(t)\,.
\end{equation}
For multi-detector in the dominant polarization basis, the coherent SNR is defined as 
\begin{equation}
	\rho_{coh}^2 = \frac{(s|F_{+}h_{0})^2 + (s|F_{+}h_{\frac{\pi}{2}})^2}{(F_{+}h_{0}|F_{+}h_{0})} + 
    \frac{(s|F_{\times}h_{0})^2 + (s|F_{\times}h_{\frac{\pi}{2}})^2}{(F_{\times}h_{0}|F_{\times}h_{0})}\,.
\end{equation}

In our work, we maximize the coherent SNR of CBC signals in a network of detectors, with and without spinning components. The coherent search filters out the signal buried in noise by maximizing it over the phase as astrophysical signals are coherent across multiple detectors. The coalescing binaries with spinning individual components causes the modulation in the profile of the overall phase of the GW waveform~\cite{KGArun2009, AB2013}. We divide the injection into three classes -- no-spin BBH injections, aligned low-spin BBH injections and aligned high-spin BBH injections. The parameters of each set are consistent with the definition in section \S\ref{subsec:InjParam}. On each of the injection sets, we use PSO to maximize the coherent SNR over the signal parameters. We set up a coherent search in a four-dimensional parameter space -- component masses and source location ($m_1, m_2, \theta, \phi$), referred to as non-spinning coherent search. We will extend the parameter space to incorporate aligned-spins extending the parameter space to six ($m_1, m_2, s_{1z}, s_{2z}, \theta, \phi$), referred to as spinning coherent search. 

\begin{figure*}
\begin{center}
\includegraphics[width=0.99\textwidth]{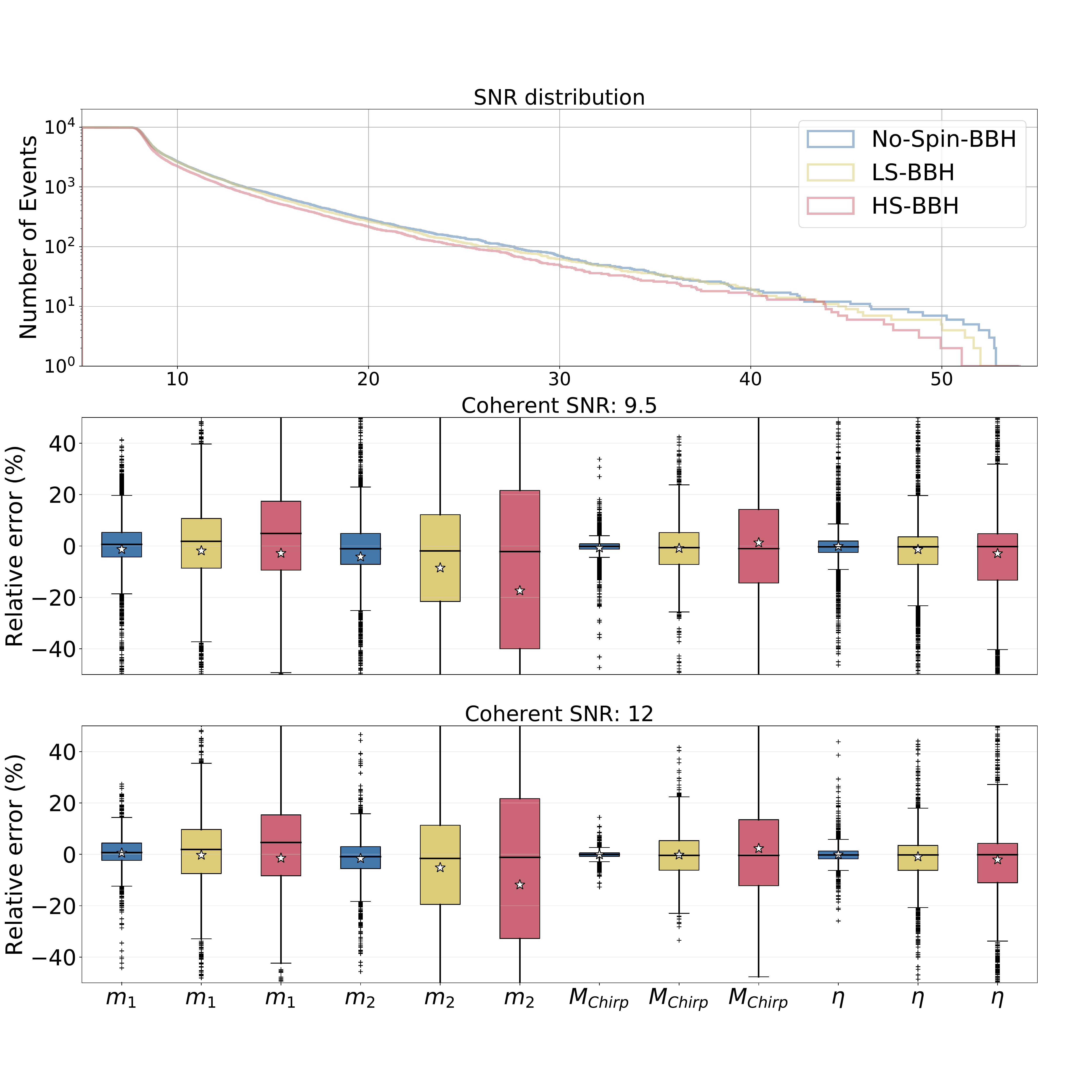}
\caption{ The plot summarizes the result of a non-spinning coherent search over three different injection set parameters -- Non-spinning BBH injections (blue), low-spin BBH injections (yellow) and high-spin BBH injections (pink). We find that coherent search is significantly affected if the injection parameters have spin. The parameters of PSO search are same for the search over each injection set. Amongst different injection sets, the component masses are the same but the spins of each component mass vary to satisfy the parameters of each class of injection set. We find the coherent search has a low error and higher detection efficiency when the injections have no spin. As the CBC spin increases the detection efficiency and estimates of CBC parameters drops. This spin-induced discrepancy is persistent with higher SNR thresholds (corresponding coherent SNR thresholds are labeled above each subplot).}
\label{fig:EffSpinCoher}
\end{center}
\end{figure*}

\subsection{Effect of Spinning Injections on Coherent search}
Coherent search maximizes over the phase overlap of the signal with the templates. The individual spin components modulate the phase of the GW waveforms, giving rise to a degeneracy between the source sky-position and component spins~\cite{KGArun2009,AB2013}, which affects the performance of any search, but especially the coherent search. We aim to study the effect of component spins of CBCs on the coherent search. To study the above we take three injection sets -- non-spinning (component masses have no spin), low-spinning and high-spinning BBH injections, each set are defined by the same set of parameters described in section \S\ref{subsec:InjParam}. We maximize the coherent SNR for each of the injections in different sets in a network of two detectors. The search is set up in a 4 dimensional parameter space (without spin {$m_1, m_2, \theta, \phi$}). We use 8 swarms with 300 particles each, totaling to 576k match filtering operations ($300 \times 8 \times 40 \times 6$) during the search. The results of the search are summarized in Fig.~\ref{fig:EffSpinCoher}. From the plot, it is evident that spinning binaries negatively impact the performance of the coherent search. We see the number of injections above the threshold (corresponding to the same FAP incoherent search) are higher when injections are non-spinning or have aligned low-spins compared to aligned high-spins injections as summarized in table \ref{tab:Ndet_Sumry}. The errors in estimated parameters from the coherent search are also lower for aligned low-spin injections as compared to aligned high-spin injections, evident from figure \ref{fig:EffSpinCoher}. One possible way to resolve this issue would be to increase the number of particles and swarms used, but we don't do so as the computational cost would increase making it less likely for the coherent search to be developed as an online search tool.

\begin{figure}
	\begin{center}
	\includegraphics[width=0.49\textwidth]{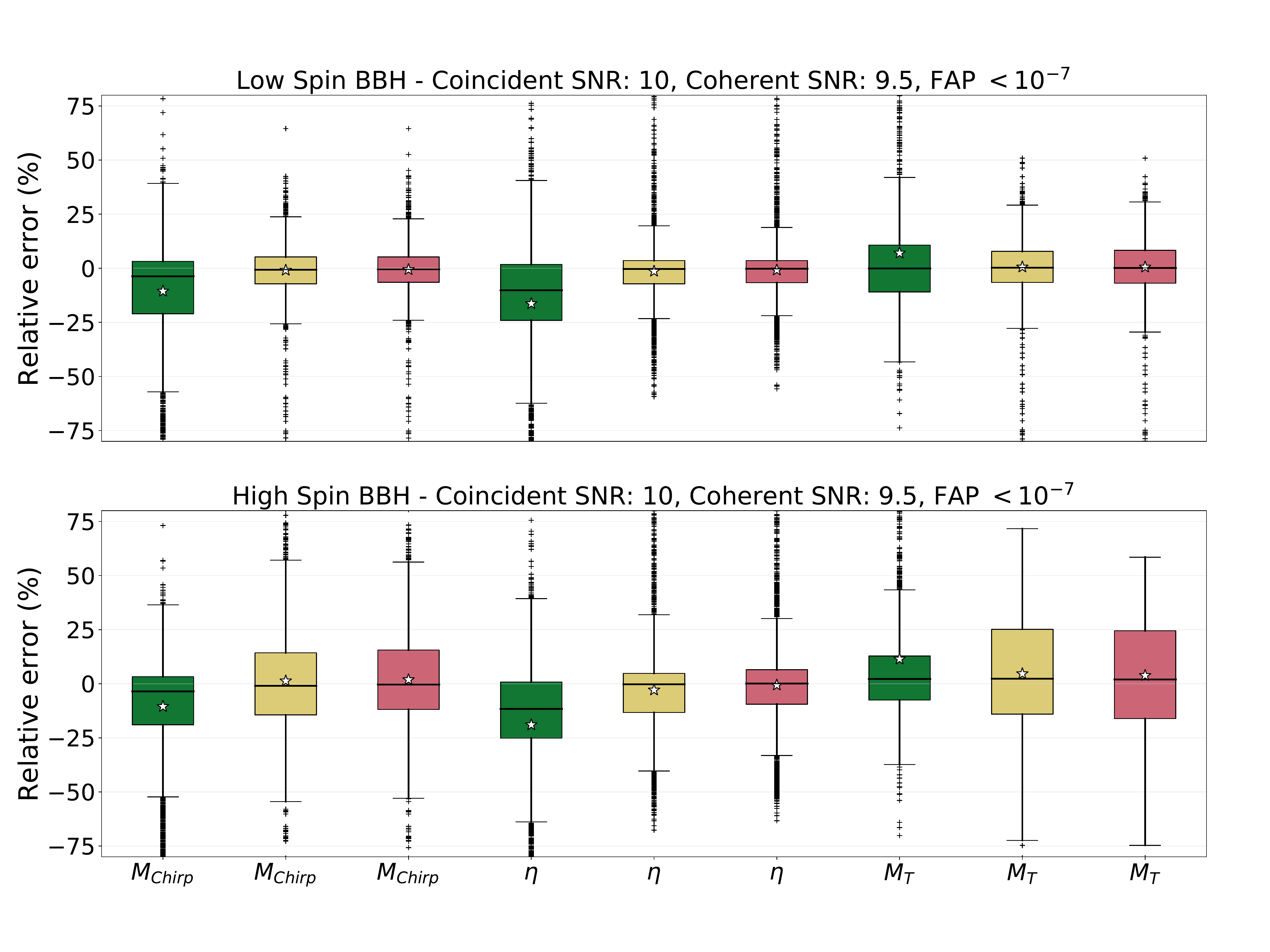}
    \caption{We compare the performance of the spinning coherent search with the template bank to standardize the comparison of the results. The coherent search is computationally more expensive than the template bank search. However, the advantage that the former offers are the ability to recover a higher number of events as summarized in table \ref{tab:BG_Th}. The plot shows for a network of two detectors HL, the error in the recovered injections (above the corresponding FAP thresholds). The two plots show the distribution for low-spin (top) and high-spin (bottom) BBH injections. We see for low-spin BBH injections the non-spinning coherent (yellow) and spinning coherent (pink) searches perform reasonably well in estimation the parameters. The errors over recovered injections are lesser when compared to the O2-template bank (green). However, for high-spin BBH injections, we see that coherent search under-performs in estimating the parameters in comparison to template bank. However, the number of injections that are recovered using coherent statistics for high-spin searches are higher than the template bank.}
	\label{fig:Coher_Smry}
	\end{center}
\end{figure}

\subsection{Extending the Parameter Space of Coherent Search: Aligned-Spin Coherent Search}
We try to resolve the problems with high-spinning injections that coherent search runs into by extending the search to an aligned-spin search -- six-dimensional space ({$m_1, m_2, s_{1z}, s_{2z}, \theta, \phi$}). We aim for detection of CBCs using coherent search, thus, we do not increase the number of points or swarms to keep the computational cost almost the same. The difference in cost of coherent search with aligned-spin and without spin is due to the generation of corresponding GW waveform templates. We use the template bank based PSO search to compare the performance of the coherent search. The results of different searches are summarized in Fig.~\ref{fig:Coher_Smry}. From the background estimation plots in section \ref{subsec:coin_bg} we observe that the background of spinning and non-spinning coherent search is almost identical. Putting the same SNR threshold on 10k BBH injections of low and high spinning system, we observe that the error estimates are similar for the two searches. However, out of the 20k BBH injections in total, the aligned-spin coherent search recovers more signals as compared to aligned-spin coincident search, as summarized in the table \ref{tab:Ndet_Sumry}. 

\begin{figure}
	\begin{center}
	\includegraphics[width=0.49\textwidth]{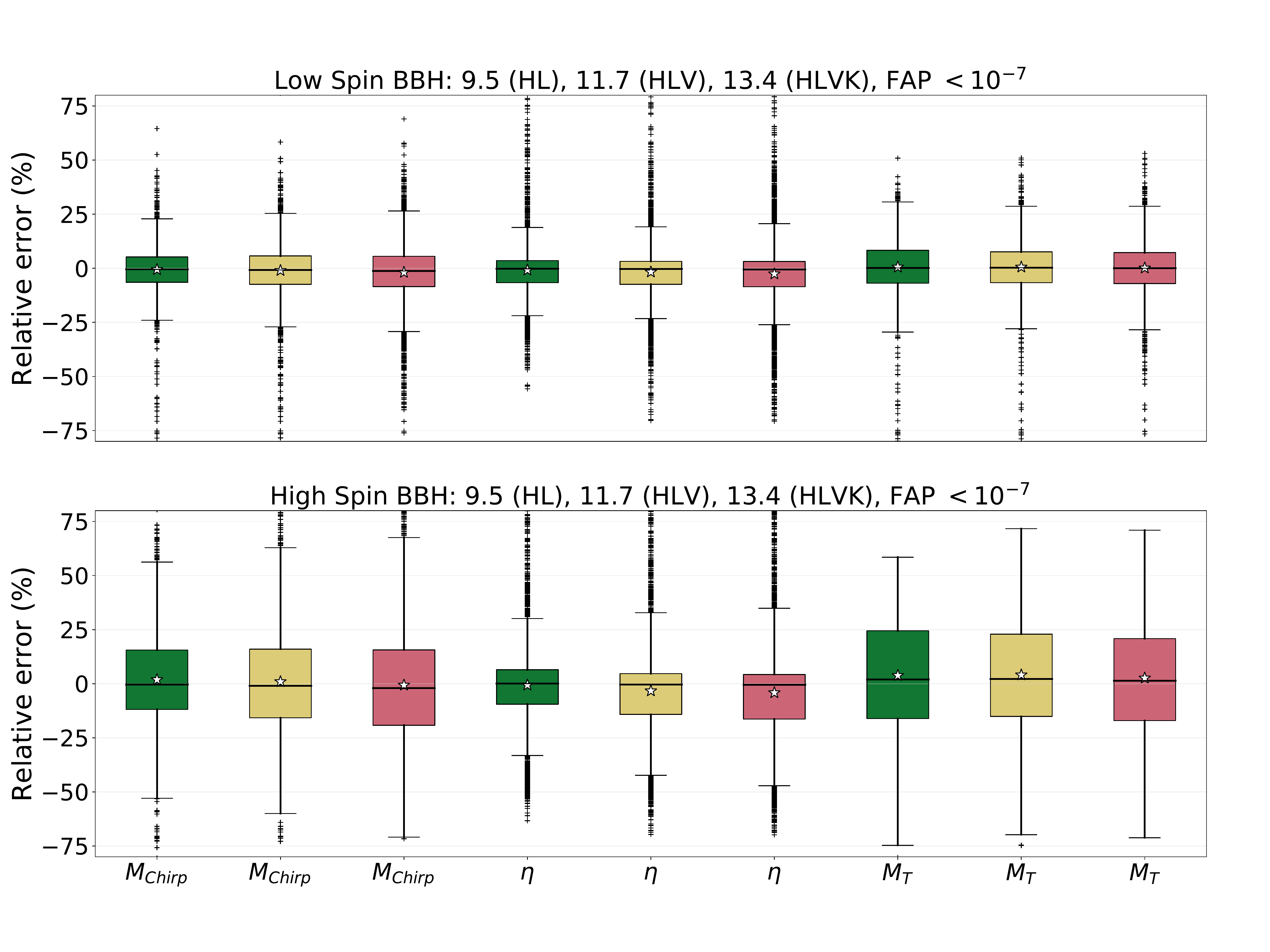}
    \caption{The plot summarizes the performance of PSO based spinning coherent searches for HL (green), HLV (yellow) and HLVK (pink) network. The top subplot shows the error distribution for low-spin BBH injections while the lower panel shows the distribution for high-spin BBH injections. The trend that coherent search performs well for CBC injections with low inherent spin as compared to high-spin CBC injections is consistent for a higher number of detectors in the network. The errors distribution spreads for a higher number of detectors because there are a higher number of events above the threshold. }
	\label{fig:Coher_Multi}
	\end{center}
\end{figure}

\begin{figure*}
\begin{center}
\includegraphics[width=0.99\textwidth]{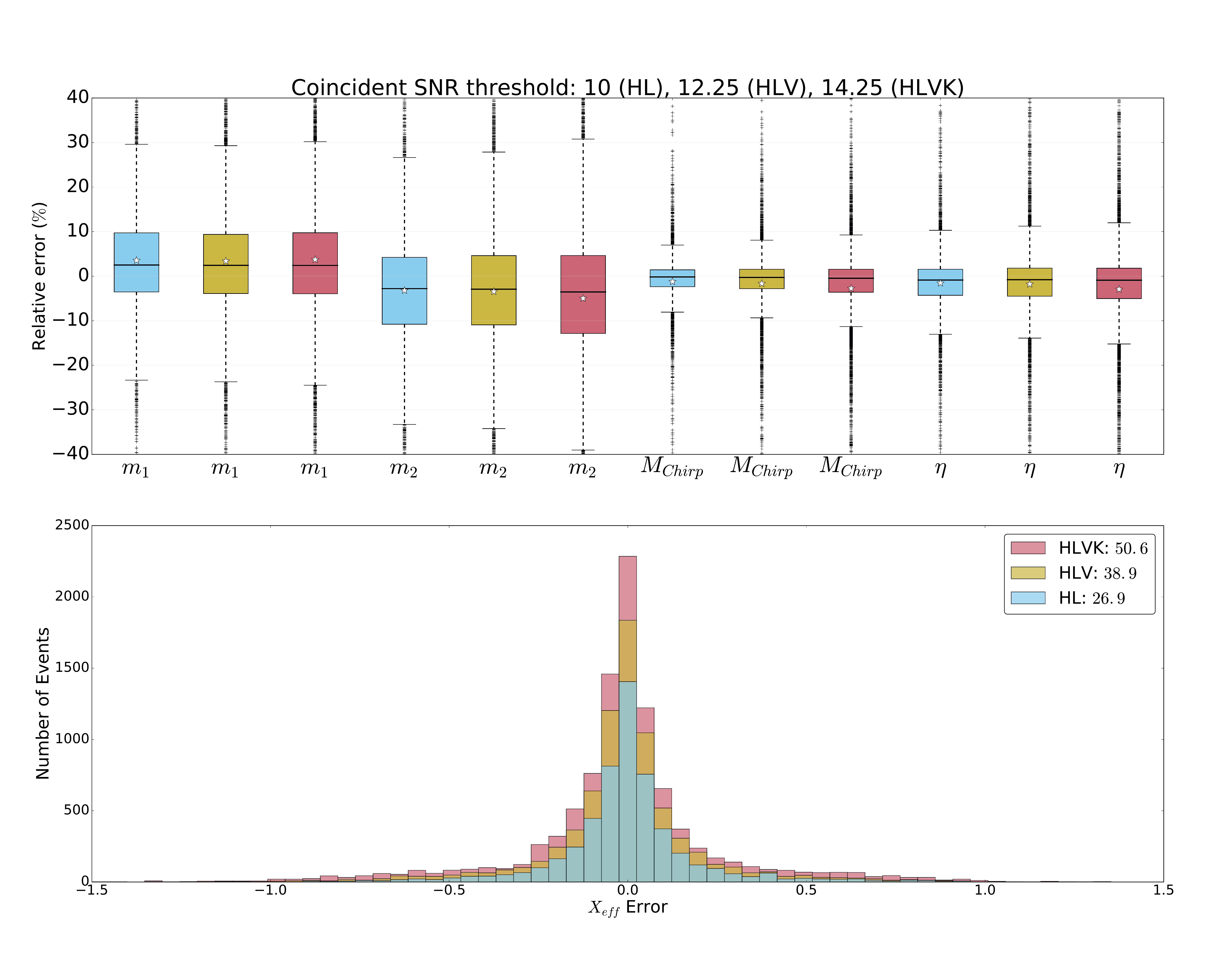}
\caption{Extending the number of detectors in the network and performing PSO based coincident search over the 20k BBH injections. The thresholds for the network are scaled by the square root of the number of detectors in the network. All possible combinations of detectors are considered in the search. That is, for HLVK detector network we consider all possible combination of two detector and three detector events that cross the corresponding thresholds. The plot above shows the error in different parameters of the CBC in HL (blue), HLV(Yellow1) and HLVK (Pink). The fraction of events recovered is indicated in the legend of the plot and in table \ref{tab:Ndet_Sumry}. The results obtained are as per the expectations with the increase in the number of detectors the fraction of recovered events increases. The distribution of error is comparable as the majority of events in higher detector networks arise from two detectors in that network. }
\label{fig:Coin_Multi}
\end{center}
\end{figure*}

\subsection{Network of detectors}\label{subsec:coin_ndet}
We can extend PSO based search to higher number of detectors in the network. We aim to recover injections with HLV and HLVK network to do consistency checks for coincident and spinning coherent searches. We see by increasing the number of detectors the number of events recovered increase, as summarized in table \ref{tab:Ndet_Sumry}. From the table, it is evident that coherent search recovers a higher number of events than coincident search and the trend is consistent with having multiple detectors in the network. The figures \ref{fig:Coher_Multi} and \ref{fig:Coin_Multi} summarize the distribution of errors in parameters of CBC for all the injections that pass the corresponding thresholds based on search and the detector network. The error estimate is almost similar for a higher number of detectors as a majority of events recovered in HLV and HLVK arise from two detectors in that network.

\begin{table*}
  	\begin{center}
  	\begin{tabular}{|c|c|c|c|c|}
  		\hline
        \ Network & Coincident Search & Coincident Search & Spinning Coherent & Spinning Coherent\\
        \ & Template Bank & PSO-based & Search & Low, High\\
        \hline
        \ HL (FAP $< 10^{-7}$) & 21.0 (10) $\%$ & 26.9 $\%$ (10) & 30.4 $\%$ (9.5) & 32.0 $\%$, 28.7 $\%$ \\
        \hline
        \ HLV (FAP $< 10^{-7}$) & - & 38.9 $\%$ (12.25) & 40.8 $\%$ (11.7) & 43.4 $\%$, 38.1 $\%$ \\
        \hline
        \ HLVK (FAP $< 10^{-7}$) & - & 50.6 $\%$ (14.25)  & 52.4 $\%$ (13.4) & 55.1 $\%$, 49.7 $\%$ \\
        \hline
  	\end{tabular}
  	\end{center}
    \caption{ The above table summarizes the fraction of events (of the 20k BBH injections) that passed the thresholds \footnote{indicated within brackets} which correspond to the same false alarm probability from background triggers in gaussian noise. The thresholds for a two detector network are obtained from Fig.~\ref{fig:BG-2Det}. For higher detectors in the network, the threshold is rescaled by the square root of the number of detectors in the network. We find that coherent search outperforms coincident search in terms of the number of events recovered and this trend is consistent with a higher number of detectors in the network. The last column shows the discrepancy in incoherent search. We see that coherent search under-performs for higher-spin injections compared to low-spin injections. However, overall coherent search seems to outperform coincident search.}
    \label{tab:Ndet_Sumry}
\end{table*}

\section{Results} \label{sec:res}
We use PSO to set up an aligned-spin coincident search, which uses 48k MFOs per detector and is an effective algorithm to search for CBCs. For a network of two detectors HL, we compare an aligned-spin PSO search with a template-bank search using the O2-template bank of Canton et al. \cite{canton2017designing}. The O2-template bank uses approximately 400k MFOs per detector compared to 48k MFOs used by aligned-spin PSO-based coincident search. As evident from the figure \ref{fig:Comp_O2} we see that at a lower computational cost, PSO recovers approximately 28 $\%$ more events as compared to template bank above the SNR of 10 and at the same time, the estimation of parameters is also better with lower error in injected parameters in the detection step as compared to the template bank. The improvement in the SNR (higher SNR achieved with PSO algorithm) in the detection stage helps BAYESTAR, a Bayesian algorithm for rapid localization, to localize the source in a smaller region in the sky. The localization capabilities of BAYESTAR is dependent on the recovered SNR in the detection stage. Chi-squared discriminator \cite{AllenChisq,Dhurandhar:2017aan} for glitches can be easily incorporated in the PSO based searches.

Next, we extend the parameter space of the search to incorporate precession. The dimensionality of the search space now includes the component masses, the orbital inclination and the component spin vectors of the two objects. To help PSO cope with the higher dimensionality of this extended parameter space, we increase the number of particles and the number of swarms used by the PSO algorithm. Using 10 swarms with 800 particles each we set up a precessing coincident search, which uses a total number of 320k MFOs per detector. The precessing PSO search recovers approximately 18 $\%$ more injections as compared to an aligned-spin PSO search with the same PSO parameters. Another striking feature of the precessing search is the accuracy in the estimation of the $\chi_{eff}$, among other parameters. Our work provides a scheme for background estimation in PSO-based searches, whether they be coincident or coherent, which can be readily applied to real data. 

In our work, we also use PSO to implement an all-sky blind coherent search. We study the effect of intrinsic spins of component masses on the performance of coherent search and find that high-spin systems affect the performance of coherent search negatively, as compared to the low-spin system. On comparing the two subgroups of injections -- aligned high-spin and aligned low-spin -- we find aligned spin coherent search recovers, for the same FAP, a higher number of events from the low-spin injection set. However, we find that in each of the two-subgroups -- high-spin and low-spin CBCs -- the coherent search outperforms coincident search by recovering a higher number of injections above the threshold corresponding to the same FAP in each search (coincident/coherent). This trend is consistent with a higher number of detectors as summarized in the table \ref{tab:Ndet_Sumry}. In conclusion, the effectiveness of PSO in a blind all-sky coherent search was demonstrated here using 576k MFOs per detector, and with parallelization techniques for the PSO algorithm \cite{schutte2004parallel,venter2006parallel}, a low-latency coherent search pipeline using PSO can be developed as well in the future.

\section{Acknowledgements} \label{sec:ack}
Special thanks are due to Bhooshan Gadre for his inputs on the manuscript and helpful discussions. We also thank Anuradha Samajdar and Steven Reyes for helpful discussions, and Jayanti Prasad for technical support. The work is partially funded by the NSF Award-1352511 (P.I. Stefan Ballmer) and the Navajbai Ratan Tata Trust. We thank Derek Davis for carefully reading the manuscript. We acknowledge the use of IUCAA LDG cluster Sarathi for the computational/numerical work.

\clearpage
\bibliographystyle{aasjournal}
\bibliography{PSO_GW}
\end{document}